\documentclass[12pt]{article}

\usepackage[a4paper, left=2.5cm, right=2.5cm, bottom=4.5cm, heightrounded]{geometry}

\usepackage[UKenglish]{babel}

\usepackage{cite}
\usepackage{comment}            % To comment several lines at once.
\usepackage{scalerel}
\usepackage{shuffle}
\usepackage{latexsym,epsfig}
\usepackage[hidelinks]{hyperref}
\usepackage{dsfont}
\usepackage{nccmath}            %fleqn and other special layouts

\usepackage{amsmath,amsfonts,amssymb,mathtools}
\usepackage{extarrows}
\usepackage{cancel, slashed}

\usepackage{upgreek}            %\uppsi
\usepackage[bbgreekl]{mathbbol} %greek letters in bb

\DeclareSymbolFontAlphabet{\mathbbg}{bbold} %Greek letters in bb
\DeclareSymbolFontAlphabet{\mathbb}{AMSb}   %Latin letters in bb

\usepackage{tikz, tikz-cd}

\usepackage{xcolor}

\numberwithin{equation}{section}
\allowdisplaybreaks
\def\bpm{\begin{pmatrix}}
\def\epm{\end{pmatrix}}
\newcommand{\defeq}{{\,\coloneq\,}}         %Definition equal
\newcommand{\eqdef}{{\,\eqcolon\,}}         %Definition equal

\newcommand{\mycomment}[1]{}
    \def\cA{{\cal A}}

    \def\cD{{\cal D}}
    \def\cE{{\cal E}}
    
    \def\cG{{\cal G}}

    \def\cK{{\cal K}}
    \def\cL{{\cal L}}
    
    \def\cN{{\cal N}}

    \def\cR{{\cal R}}

    \def\cX{{\cal X}}

    \newcommand{\bbLambda}{\mathbbg{\Lambda}}   %Gauge parameters
    \newcommand{\bbH}{\mathbb{H}}               %Fields
    \newcommand{\bbE}{\mathbb{E}}               %Equations of Motion
    \newcommand{\bbN}{\mathbb{N}}               %Noether Identities
    \newcommand{\bbB}{\mathbb{B}}               %Operator B
    \newcommand{\bbSigma}{\mathbbg{\Sigma}}     %Operator Sigma
    \newcommand{\1}{\mathbbg{1}}

    \newcommand{\bfXi}{\mathbf{\Xi}}    

    \newcommand{\talpha}{{\tilde\alpha}}
    \newcommand{\tbeta}{{\tilde\beta}}

    \newcommand{\tmu}{{\tilde\mu}}
    \newcommand{\tnu}{{\tilde\nu}}
    \newcommand{\trho}{{\tilde\rho}}
    
    \newcommand{\tlambda}{{\tilde\lambda}}
    \newcommand{\tepsilon}{{\tilde\epsilon}}
    \newcommand{\tpsi}{{\tilde\psi}}
    \newcommand{\te}{{\tilde e}}
    \newcommand{\tdel}{{\tilde\partial}}

    \def\del{\partial}
    \def\Box{\square\,}
    \def\delslash{\slashed{\del}}

    \def\vbar{\,|\,}
    \def\cdott{\!\cdot\!}
    
    \def\Tr{{\rm Tr}}

\begin{document}

\begin{flushright}
    August 2025 \\
    HU-EP-25/29-RTG
\end{flushright}

\vskip 13mm

\begin{center}
    
    {\large\textbf{The Double Copy of Maximal Supersymmetry in $D=10$}}
    
    \vskip 10mm

    Roberto Bonezzi$^1$, Giuseppe Casale$^2$, Olaf Hohm$^3$

    \vskip 10mm

    {\small\textit{Institut f\"ur Physik, Humboldt-Universit\"at zu Berlin}} \\
    {\small\textit{Zum Großen Windkanal 2, D-12489 Berlin, Germany}}

    \vskip 25mm

    {\bf{Abstract}}

\end{center}

\begin{quote}
We continue the program of using  homotopy algebras to obtain off-shell, local and gauge redundant derivations of the double copy relations between  gauge theory and gravity. We apply it to ${\cal N}=1$ super-Yang-Mills theory in $D=10$ in order to obtain type IIA or type IIB supergravity, at least to cubic order in fields. Furthermore, we show how the super-Lie algebra of global supersymmetries, acting on the homotopy algebra of ${\cal N}=1$ super-Yang-Mills theory,  double copies to the maximal supersymmetry of supergravity. 
\end{quote}

\vfill

\noindent
{\small
$^1$ bonezzi@physik.hu-berlin.de \\
$^2$ giuseppe.casale@physik.hu-berlin.de \\
$^3$ ohohm@physik.hu-berlin.de
}

\thispagestyle{empty}

\newpage

\clearpage
\pagenumbering{arabic}

\tableofcontents

\setcounter{footnote}{0}

%%%%%%%%%%%%%%%%%%%%%%%%%%%%%%%%%%
%%%%%%%%%%%%%%%%%%%%%%%%%%%%%%%%
%% BEGIN INTRODUCTION %%%%%%%%%%%%
%%%%%%%%%%%%%%%%%%%%%%%%%%%%%%%%
%%%%%%%%%%%%%%%%%%%%%%%%%%%%%%%%%%
\section{Introduction}
\label{Sec:Introduction}
The double copy denotes a deep relation between gauge theories like (super-)Yang-Mills theory and gravity. 
It was originally discovered in the realm of scattering amplitudes \cite{Bern:2008qj, Bern:2010ue, Bern:2010yg, Bjerrum-Bohr:2010pnr,Mafra:2011kj, Carrasco:2011mn,Bern:2011rj, Bern:2019prr, Bern:2022wqg, Adamo:2022dcm}, based on the earlier Kawai-Lewellen-Tye (KLT) relations between open and closed string scattering amplitudes \cite{Kawai:1985xq}. The arguably most intriguing double copy relation is between the UV finite ${\cal N}=4$ super-Yang-Mills theory in $D=4$  and ${\cal N}=8$ supergravity, which was used to show that the latter has an improved  UV-finite  behavior \cite{Bern:2012uf, Bern:2017yxu, Bern:2017ucb}. 
However, the double copy is also applicable to pure Yang-Mills theory, where it yields `${\cal N}=0$ supergravity' -- Einstein gravity coupled to a two-form (B-field) and a scalar (dilaton) --, which in turn is naturally formulated as a double field theory (DFT)  \cite{Siegel:1993th,Hull:2009mi,Hohm:2010jy,Hohm:2010pp}. 

While the double copy is well understood at the level of tree-level scattering amplitudes, it remains as an open problem to obtain a real first-principle derivation, say starting from the off-shell and gauge redundant Lagrangian, as would be needed, for instance,  in order to double copy classical solutions. 
In recent years a powerful new approach has emerged that utilizes the formulation of quantum field theory (QFT) in terms of homotopy algebras \cite{Zwiebach:1992ie, Lada:1992wc, Hohm:2017pnh, Jurco:2018sby}. 
Specifically, this framework allows one to `color-strip' Yang-Mills theory and to uncover a hidden `kinematic' homotopy algebra, introduced and named BV$_{\infty}^{\square}$ by Reiterer \cite{Reiterer:2019dys}. See  \cite{Lian:1992mn, Zeitlin:2009tj, Zeitlin:2014xma} for earlier work and \cite{Borsten:2021hua,Ben-Shahar:2021doh, Diaz-Jaramillo:2021wtl, Ben-Shahar:2021zww, Bonezzi:2022yuh,Borsten:2022vtg,Borsten:2022ouu, Bonezzi:2022bse, Bonezzi:2023pox,Borsten:2023ned,  Borsten:2023paw,Bonezzi:2023lkx,Bonezzi:2024dlv, Ben-Shahar:2024dju, Bonezzi:2025anl} for the recent developments\footnote{See also \cite{Anastasiou:2018rdx, Borsten:2020xbt, Borsten:2020zgj, Godazgar:2022gfw} for related approaches to off-shell double copy in the BRST framework.}. 

Recently, we continued this program by applying it to ${\cal N}=4$ super Yang-Mills theory in four dimensions \cite{Bonezzi:2025anl}, and we showed, to cubic order in fields, that the double copy indeed yields ${\cal N}=8$ supergravity in a DFT formulation. 
Here it is important to recognize that the homotopy algebra encodes  the \textit{perturbative} formulation, with the action expanded around a background, which we take to be flat space, to yield quadratic, cubic and higher order terms. 
The local gauge symmetries of supergravity then split into two kinds of symmetries: \textit{global} spacetime (super-)isometries that leave the background invariant, and \textit{local} diffeomorphisms and supersymmetry transformations. 
While the latter are naturally part of the homotopy algebra, the global symmetries are more subtle and must be viewed as an external action by the super-Lie algebra on the homotopy algebra. See also \cite{Anastasiou:2015vba} for a discussion of doubling global symmetries. 
In  \cite{Bonezzi:2025anl} we proved that the ${\cal N}=4$ supersymmetry then indeed implies, up to and including cubic order in fields,  
the doubled ${\cal N}=8$ supersymmetry of maximal supergravity in four dimensions. More precisely, the double copy map via homotopy algebras has been established to quartic order in  \cite{Bonezzi:2022bse}, but the proof that the global supersymmetry double copies so far is restricted to the cubic theory. 

In this paper our goal is two-fold. First, we want to extend the construction of \cite{Bonezzi:2025anl} to super-Yang-Mills theory in $D=10$. This case is interesting since $D=10$ is the dimension of superstring theory, and there are two maximally supersymmetric theories: the type IIA and type IIB theory. 
Second, we want to push further the interpretation of global symmetries as external structures acting on the homotopy algebra by showing how also \textit{closure} of the supersymmetry algebra, be it on-shell or off-shell, follows purely from the double copy perspective. 

The remainder of this paper is organized as follows. In sec.~2 we introduce the kinematic algebra of ${\cal N}=1$ super-Yang-Mills theory in $D=10$ and work out the action of the super-Poincar\'e algebra in detail. 
In sec.~3 we then turn to the double copy and prove, at the level of the free theory, that it yields either type IIA or type IIB supergravity. 
Moreover, we analyze the double copy of the global (super-)symmetries in detail. 
Together with the match of the free theories, this implies that to cubic order 
we indeed obtain either type IIA or type IIB supergravity. 
We close in sec.~4 with a conclusion section, while our spinor conventions are summarized in an appendix.

%%%%%%%%%%%%%%%%%%%%%%%%%%%%%%%%%%
%%%%%%%%%%%%%%%%%%%%%%%%%%%%%%%%
%% BEGIN SECTION 3 %%%%%%%%%%%%%%%
%%%%%%%%%%%%%%%%%%%%%%%%%%%%%%%%
%%%%%%%%%%%%%%%%%%%%%%%%%%%%%%%%%%
\section{Kinematic Algebra of $D=10$ Super Yang-Mills}
\label{Sec:10DKinAlg}

In this section we provide the action, the equations of motion and the (super)symmetry transformations for super Yang-Mills theory in 10 dimensions. Thereafter, the perturbative theory will be reformulated algebraically in terms of an $L_\infty$ algebra, a homotopy-extension of a Lie algebra underlying the perturbative structure of any gauge theory \cite{Hohm:2017pnh}. This language prepares us for the next step towards the double copy: color-stripping. Stripping the color degrees of freedom off a theory in this formalism entails going from an $L_\infty$ algebra to a $C_\infty$ algebra. This results in the \emph{kinematic algebra} of super Yang-Mills, which is the fundamental building block underlying the double-copy to a supergravity theory, as we will show in Section \ref{Sec:DC_Sugra}.

%%%%%%%%%%%%%%%%%%%%%%%%%%%%%%%%
%%%%%%%%%%%%%%%%%%%%%%%%%%%%%%
%%%%%%%%%%%%%%%%%%%%%%%%%%%%%%%%
\subsection{Action and Symmetries}
We start from the action of 10-dimensional super Yang-Mills theory:
\begin{equation}\
S_{\mathrm{SYM}} = \int d^{10}x\; \Tr\left[ -\frac14\, F_{\mu\nu}F^{\mu\nu} + \frac i2\, \chi\gamma^\mu\cD_\mu \chi \,\right]\;,
    \label{Action:SYM_full}
\end{equation}
where $\mu,\nu = 0,\dots,9$ are spacetime Lorentz indices, $F_{\mu\nu} = 2\,\del_{[\mu}A_{\nu]} + [A_\mu,A_\nu]$ is the field strength associated with the gauge field $A_\mu$, $\cD_\mu \defeq \del_\mu + [A_\mu,-]$ is the covariant derivative, and finally $\chi^\alpha$ is a real Majorana-Weyl spinor -- the so-called \emph{gaugino} -- endowed with a spinor index $\alpha=1,\dots,16$. Both the gauge field and the gaugino are Lie algebra-valued: $A_\mu=A_\mu^a\,t_a$ and $\chi^\alpha=\chi^{\alpha a}\,t_a$, where $t_a\in\mathfrak{g}$ are generators of the color Lie algebra.
The matrices $(\gamma^\mu)_{\alpha\beta}$, $(\bar\gamma^\mu)^{\alpha\beta}$ are a set of generalized Pauli matrices obeying the usual Clifford-like relation
\begin{equation}
    (\bar\gamma^\mu)^{\alpha\lambda}(\gamma^\nu)_{\lambda\beta} + (\bar\gamma^\nu)^{\alpha\lambda}(\gamma^\mu)_{\lambda\beta} = 2\,\eta^{\mu\nu}\delta^\alpha_\beta \;,
\end{equation}
$\eta^{\mu\nu}$ being the flat Minkowski metric. In the following we will omit the bar on the generalized Pauli matrices $\bar\gamma^\mu$ with upper spinor indices, as the difference between the two sets of matrices $\gamma^\mu$ and $\bar\gamma^\mu$ will be understood from the context.

The action \eqref{Action:SYM_full} is invariant under the (local) gauge transformations
\begin{align}
    \delta_\Lambda A_\mu = \cD_\mu\Lambda\;, \quad\quad \delta_\Lambda \chi = [\chi,\Lambda]\;, 
\end{align}
as well as under global super-Poincaré symmetries. These consist  of Poincaré transformations
\begin{align}
    \delta_\xi A_\mu = \xi^\rho\del_\rho A_\mu + \del_\mu\xi^\rho A_\rho\;, \quad\quad \delta_\xi \chi = \xi^\mu\del_\mu\chi + \tfrac14\,\omega^{\mu\nu}\gamma_{\mu\nu}\chi\;, 
    \label{Eq:Diffeos_fields}
\end{align}
generated by the Killing vector $\xi^\mu \defeq a^\mu - \omega^\mu{}_\nu x^\nu$, with constant $a^\mu$ and $\omega_{\mu\nu} = -\omega_{\nu\mu}$, and global supersymmetry transformations
generated by the fermionic parameter $\epsilon^\alpha$, 
\begin{align}
    \delta_\epsilon A_\mu = i\,\epsilon\gamma_\mu\chi\;, \quad\quad \delta_\epsilon \chi = \frac12\,F_{\mu\nu}\gamma^{\mu\nu}\epsilon\;,
    \label{Eq:GlobalSUSY_fields}
\end{align}
where the Lorentz spin generator acting on Majorana-Weyl spinors with upper indices is defined by
\begin{equation}
    (\gamma^{\mu\nu})^\alpha{}_\beta \coloneq \frac12 \Big[ (\bar\gamma^\mu)^{\alpha\lambda}(\gamma^\nu)_{\lambda\beta} - (\bar\gamma^\nu)^{\alpha\lambda}(\gamma^\mu)_{\lambda\beta} \Big] \;.
\end{equation}
We refer to Appendix \ref{App:Conventions} and to the works \cite{GuillenQuiroz:2016sla, Guillen:2019pnz} for more details on the definition and properties of generalized Pauli matrices, as well as on generalities on spinors in $D=10$.
Finally, the equations of motion resulting from the action \eqref{Action:SYM_full} are:
\begin{equation}\begin{gathered}
    \cD_\nu F^{\nu\mu} - \frac i2\, [\chi, \gamma^\mu\chi] = 0\;, \quad\quad
    i\,\gamma^\mu\cD_\mu\chi = 0\;.
\end{gathered}\end{equation}

As pointed out in \cite{Bonezzi:2022yuh}, in view of performing a local double copy in terms of unconstrained fields, it is beneficial to rephrase the free part of the gauge theory by introducing an auxiliary field $\varphi$, obeying the on-shell condition
\begin{equation}
    \varphi=\del\!\cdot\!A\;, 
\end{equation}
so that the action gets modified to
\begin{equation}\begin{split}
    S_{\mathrm{SYM}} = \int d^{10}x\; \Tr\,&\left[ \, \frac12 A_\mu\Box A^\mu + (\del\!\cdot\!A)\varphi - \frac12\,\varphi^2 + \frac i2\,\chi\gamma^\mu\del_\mu\chi \right. \\
    &- \del_\mu A_\nu[A^\mu,A^\nu] + \frac i2\, \chi\gamma^\mu[A_\mu,\chi] \\
    &- \left.\frac14[A_\mu,A_\nu][A^\mu,A^\nu] \, \right]\;. 
    \label{Action:SYM_full_explicit}
\end{split}\end{equation}
For later convenience here we have written out  the quadratic, cubic and quartic terms respectively on the first, second and third lines. Upon integrating out the auxiliary scalar $\varphi$, the action reduces to the standard form \eqref{Action:SYM_full}.

%%%%%%%%%%%%%%%%%%%%%%%%%%%%%%%%
%%%%%%%%%%%%%%%%%%%%%%%%%%%%%%
%%%%%%%%%%%%%%%%%%%%%%%%%%%%%%%%
\subsection{Homotopy Algebra Formulations}
\label{Subsec:SC_Hom_Form}
Let us now reformulate super Yang-Mills theory within the homotopy algebra framework following the prescription explained in \cite{Bonezzi:2025anl}. In particular, we will first provide a description in terms of an $L_\infty$ algebra. Subsequently, we will remove the color from the theory. As anticipated, this operation results in a description of the theory based on a $C_\infty$ algebra \cite{Zeitlin:2008cc,Borsten:2021hua,Bonezzi:2022yuh} endowed with some additional operations, which will be discussed in detail in the following. Such a formulation allows us to identify a sector of the so-called kinematic algebra of the gauge theory, which is a homotopy generalization of Batalin-Vilkovisky (BV) algebras \cite{Reiterer:2019dys,Bonezzi:2022bse}.

\subsubsection*{$L_\infty$ Formulation}

In general, an $L_\infty$ algebra $(\cX,B_n)$ is defined on a graded vector space
\begin{equation}
    \cX = \bigoplus_{i\in\mathbb{Z}} X_i
\end{equation}
and is equipped with a (potentially infinite) set of graded-symmetric $n$-linear maps (also known as $n$-\emph{brackets})
\begin{equation}
    B_n \;:\; \cX^{\otimes n} \longrightarrow \cX\;, \quad |B_n|=1\;, \quad n\geq1\;, 
\end{equation}
which encode  the perturbative expansion of the field theory, as every $B_n$ provides information about interactions at order $n+1$. Indeed, after introducing a suitable inner product $\langle\cdot\,,\cdot\rangle$ the perturbative action can be expressed as
\begin{equation}
    S = \sum_{n=1}^{\infty}\frac1{(n+1)!}\langle\cA,B_n(\underbrace{\cA,\dots,\cA}_{\text{$n$ times}})\rangle \;,
    \label{Action:L_infty}
\end{equation}
where $\cA$ denotes generic fields in degree zero, so that $B_n(\cA,\dots,\cA)$ is totally symmetric in the inputs.
Consequently, the equations of motion may be written in the generalized Maurer-Cartan form
\begin{equation}\label{MC equation}
    \frac{\delta S}{\delta\cA} = \sum_{n=1}^{\infty}\frac1{n!} B_n(\underbrace{\cA,\dots,\cA}_{\text{$n$ times}}) \;.
\end{equation}
Additionally, the $n$-brackets obey some generalized Jacobi identities, together with nilpotency of the first map: $B_1^2=0$. In particular, this last property allows us to reorganize the theory in a chain complex as follows:
\begin{equation}
   \begin{tikzcd}[row sep=0pt, column sep={2cm,between origins}]
        \cdots \arrow[r,"B_1"] & X_{-1} \arrow[r,"B_1"] & X_{0} \arrow[r,"B_1"]  &X_{1}\arrow[r,"B_1"] & X_{2}\arrow[r,"B_1"] & \cdots \\
        & \Uplambda & \cA & \cE & \cN &
   \end{tikzcd}
\end{equation}
where $\Uplambda$, $\cA$, $\cE$ and $\cN$ are respectively the sets of gauge parameters, fields, equations of motions and Noether identities.

In our case, the chain complex decomposes into a bosonic and a fermionic chain complexes: 
\begin{equation}
    \cX_{\rm SYM} = \cX^{\rm B} \oplus \cX^{\rm F} \;.
\end{equation}
The bosonic chain complex reads
\begin{equation}
\label{CC:SYM_Linf_bos}
   \begin{tikzcd}[row sep=0pt, column sep={2.5cm,between origins}]
        X^{\rm B}_{-1} \arrow[r,"B_1"] & X^{\rm B}_{0} \arrow[r,"B_1"]  &X^{\rm B}_{1}\arrow[r,"B_1"] &X^{\rm B}_{2} \\
        \Lambda & A_\mu & E & \\
        & \varphi & E_\mu & N
   \end{tikzcd}
\end{equation}
where $E$ and $E_\mu$ represent  the bosonic equations of motion for $A_\mu$ and $\varphi$, and $N$ represents the bosonic Noether identity. The action of $B_1$ on the gauge parameter $\Lambda$ provides the linearized gauge variations of the fields
\begin{equation}
    B_1(\Lambda) = \bpm \del_\mu\Lambda \\[1mm] \Box\Lambda \epm \,\in X_0^{\rm B} \;,  
\end{equation}
that is to say $\delta_\Lambda^{\rm lin}\cA = B_1(\Lambda)$. On fields, instead, the action
\begin{equation}
    B_1 \bpm A_\mu \\[1mm] \varphi \epm =
    \bpm \del\cdott A - \varphi \\[1mm] \Box A_\mu - \del_\mu\varphi \epm \, \in\,X_1^{\rm B}
\end{equation}
gives the bosonic linearized equations of motion $B_1(\cA_{\rm B}) = 0$, $\cA_{\rm B}$ being the set of all bosonic fields. Finally, acting on the equations of motion
\begin{equation}
    B_1 \bpm E \\[1mm] E^\mu \epm =
    \bpm \varnothing \\[1mm] \Box E - \del_\mu E^\mu \epm \, \in\,X_2^{\rm B}
\end{equation}
one obtains the Noether identity as $B_1(\cE_{\rm B}) = 0$ when $\cE_{\rm B}=B_1(\cA_{\rm B})$ is  the bosonic equation of motion.

The structure of the bosonic complex \eqref{CC:SYM_Linf_bos} allows us to introduce a further nilpotent operator $b$ of degree $\vbar b \vbar=-1$ obeying the defining properties
\begin{equation}\label{b_defining_Linf}
    b^2=0\;, \qquad b\,B_1+B_1\,b=\Box \;.
\end{equation}
Such operator acts as a \emph{degree shift} on the objects of the bosonic chain complex, corresponding to an isomorphism between the first and the second line of the complex:
\begin{equation}
\label{CC:SYM_Linf_bos_b}
   \begin{tikzcd}[row sep=0pt, column sep={2.5cm,between origins}]
        X^{\rm B}_{-1} \arrow[r,"B_1"] & X^{\rm B}_{0} \arrow[r,"B_1"]  &X^{\rm B}_{1}\arrow[r,"B_1"] &X^{\rm B}_{2} \\
        \Lambda & A_\mu & E & \\
        & \arrow[ul,"b"]\varphi & \arrow[ul,"b"]E_\mu & \arrow[ul,"b"]N
   \end{tikzcd} \;.
\end{equation}
The explicit action of the operator $b$ on the bosonic elements of the complex is given by
\begin{subequations}\begin{align}
    b\bpm \varnothing \\[1mm] N \epm &= \bpm N \\[1mm] 0 \epm \,\in\, X_1^{\rm B} \;, \\[2mm]
    b\bpm E \\[1mm] E_\mu \epm &= \bpm E_\mu \\[1mm] 0 \epm \,\in\, X_0^{\rm B}\;, \\[2mm]
    b\bpm A_\mu \\[1mm] \varphi \epm &= \bpm \varphi \\[1mm] \varnothing \epm \,\in\, X_{-1}^{\rm B}\;.
\end{align}\end{subequations}
This operator is a crucial ingredient of the kinematic algebra. Its simple action as a degree shift, which is the rationale for introducing the auxiliary $\varphi$, facilitates the construction of a \emph{local} double copy theory based on this gauge theory. 

As for the fermionic chain complex,  a standard formulation of $D=10$ SYM would suggest including only the spaces of spinor fields and related  equations of motion:
\begin{equation}
   \begin{tikzcd}[row sep=0pt, column sep={2.5cm,between origins}]
        0 \arrow[r] & X^{\rm F}_{0} \arrow[r,"B_1"]  &X^{\rm F}_{1}\arrow[r] &0 \\
        & \chi & \cE_\chi &
   \end{tikzcd}
\end{equation}
with the only non-vanishing $B_1(\chi)=i\gamma^\mu\del_\mu\chi\in X_1^{\rm F}$. On the other hand, a fermionic complex like this would suggest to define an operator $b$, still fulfilling \eqref{b_defining_Linf}, as 
\begin{equation}\label{b_naive}
    b(\cE_\chi) = -i\,\gamma^\mu\del_\mu \cE_\chi\,\in\,X_0^{\rm F}\;,\quad b(\chi) = 0\;,
\end{equation}
but this would lead to a double copy theory whose fields would be subject to undesirable differential constraints. A possible solution to this problem consists in conveniently redefining the action of the differential $B_1$ on the gaugino $\chi$ so that the space of fermionic equations of motion includes a second \emph{redundant} equation, which of course does not change the dynamics. This strategy defines a doublet of spinors in the subspace $X_1^{\rm F}$ given by
\begin{equation}\label{B1_ferm}
    B_1(\chi) = \bpm i\,\gamma^\mu\del_\mu\chi \\[1mm] \Box\chi \epm\,\in\,X_1^{\rm F}
\end{equation}
and allows the operator $b$ to be defined as a degree shift also on the fermionic chain complex:
\begin{equation}\label{b_ferm}
    b \bpm \cE_\chi \\[1mm] E_\chi \epm = E_\chi\,\in\,X_0^{\rm F}\;, \quad b(\chi)=0\;.
\end{equation}
The drawback of this procedure is that \eqref{B1_ferm} requires the fermionic chain complex to be semi-infinite, that is, to acquire a semi-infinite number of cascading (redundant) Noether identities caused by the presence of two dependent equations of motion, as schematically shown in the following: 
\begin{equation}
\label{CC:SYM_Linf_fer}
   \begin{tikzcd}[row sep=10pt, column sep={2.5cm,between origins}]
        0 \arrow[r] & X^{\rm F}_{0} \arrow[r,"B_1"]  &X^{\rm F}_{1}\arrow[r,"B_1"] &X^{\rm F}_{2}\arrow[r,"B_1"] & \cdots \\
        & \chi \arrow[r,"i\,\gamma^\mu\del_\mu"]\arrow[dr,"\Box"] & \cE_\chi \arrow[r,"i\,\gamma^\nu\del_\nu"] & \cN_\chi \arrow[r,"i\,\gamma^\rho\del_\rho"]\arrow[dr,"\Box"] & \cdots \\
        & & E_\chi \arrow[ru,"1"]\arrow[r,"-i\,\gamma^\mu\del_\mu"] & N_\chi \arrow[r,"-i\,\gamma^\nu\del_\nu"] & \cdots
   \end{tikzcd}\quad.
\end{equation}
Notwithstanding, the resulting formulation allows for the introduction of the $b$ operator as a degree shift operator acting uniformly on the complete chain complex
\begin{equation}
\label{CC:SYM_Linf_full}
   \begin{tikzcd}[row sep=0pt, column sep={2.5cm,between origins}]
        X_{-1} \arrow[r,"B_1"] & X_{0} \arrow[r,"B_1"]  &X_{1}\arrow[r,"B_1"] &X_{2}\arrow[r,"B_1"] & \cdots \\
        \Lambda & A_\mu, \chi & E, \cE_\chi & \cN_\chi & \cdots \\
        & \arrow[ul,"b"]\varphi & \arrow[ul,"b"]E_\mu, E_\chi & \arrow[ul,"b"]N, N_\chi & \cdots
   \end{tikzcd}
\end{equation}
which bears a local double copy written by means of unconstrained fields. 

As for the nonlinear structure of the theory, here we will not present explicitly the brackets $B_2$, $B_3$ and their relations, as they have been discussed in detail in \cite{Bonezzi:2025anl} for the case of super Yang-Mills theory in four dimensions.

\subsubsection*{Kinematic Algebra}

The next step towards the double copy of 10-dimensional super Yang-Mills is to write down its kinematic algebra. This translates into a $C_\infty$ algebra $(\cK_{\rm SYM},m_n)$, defined on a graded vector space
\begin{equation}
    \cK_{\rm SYM} = \bigoplus_{i=0}^\infty K_i
\end{equation}
and endowed with a set of $n$-linear maps
\begin{equation}
    m_n \;:\; \cK^{\otimes n} \longrightarrow \cK \;, \quad |m_n| = 2-n \;, \quad n\ge 1 \;,
\end{equation}
generalizing the product of a commutative, associative algebra.
The relations between the $n$-brackets defined on the whole $L_\infty$ algebra given by
\begin{equation}
    \cX_{\rm SYM} = \cK_{\rm SYM} \otimes \mathfrak{g}\;, \quad \mathfrak{g} = \text{color Lie algebra}\;, 
\end{equation}
and the maps $m_n$ of the color-stripped algebra $\cK_{\rm SYM}$ are expressed by the formulas
\begin{fleqn}
\begin{subequations}\begin{align}
    B_1(x) &= m_1(u^a)\otimes t_a \;, \\[2mm]
    B_2(x_1,x_2) &= (-1)^{x_1}f^a{}_{bc}\,m_2(u_1^b,u_2^c)\otimes t_a \;, \\[1mm]
    B_3(x_1,x_2,x_3) &= f^a{}_{be}f^e{}_{cd}\,\Big[(-1)^{x_2}m_3(u_{1}^b,u_2^c,u_{3}^d)\! +\! (-1)^{x_1(x_2+1)}m_3(u_{2}^b,u_1^c,u_{3}^d)\Big]\! \otimes t_a \,,
\end{align}\label{Eq:Linf_Cinf_relations}\end{subequations}
\end{fleqn}
where $t_a$ and $f^a{}_{bc}$ are respectively generators and structure constants of $\mathfrak{g}$, while $x_i = u_i^a\otimes t_a \in \cX_{\rm SYM}$ are arbitrary elements of degree $|x_i| = |u_i| -1$ of the full $L_\infty$ algebra and $u_i\in\cK_{\rm SYM}$ are their $C_\infty$ counterparts. The color-stripped chain complex $(\cK_{\rm SYM}, m_1)$ can also be factorized into its bosonic and fermionic sectors $\cK_{\rm SYM}=\cK^{\rm B}\oplus\cK^{\rm F}$, together yielding the semi-infinite complex
\begin{equation}
\label{CC:SYM_Cinf_full}
   \begin{tikzcd}[row sep=0pt, column sep={2.5cm,between origins}]
        K_{0} \arrow[r,"m_1"] & K_{1} \arrow[r,"m_1"]  &K_{2}\arrow[r,"m_1"] &K_{3}\arrow[r,"m_1"] & \cdots \\
        \Lambda & A_\mu, \chi & E, \cE_\chi & \cN_\chi & \cdots \\
        & \varphi & E_\mu, E_\chi & N, N_\chi & \cdots
   \end{tikzcd}
\end{equation}
where we use the same symbols as in \eqref{CC:SYM_Linf_full} to denote the color-stripped elements.
Analogously to the $L_\infty$ case, also in the $C_\infty$ case we can introduce an operator $b$ acting as degree shift on the elements of the chain complex and obeying
\begin{equation}\label{b_defining_Cinf}
    b^2=0\;, \qquad b\,m_1+m_1\, b=\Box \;, \quad \vbar b \vbar = -1 \;.
\end{equation}
The fact that the relations \eqref{b_defining_Linf} and \eqref{b_defining_Cinf} have the same form shows that $b$ does not act on the color algebra, but rather only on the kinematic algebra, exactly as $B_1$. The presence of the extra operator $b$ itself enriches the $C_\infty$ algebra and suggests that the  kinematic algebra at the heart  of the double copy is a much larger one, which goes under the name of ${\rm BV}_\infty^\Box$ \cite{Reiterer:2019dys}. This  is a homotopy generalization of Batalin-Vilkovisky ($\rm BV$) algebras, that includes entire hierarchies of higher products, brackets, etc.

%%%%%%%%%%%%%%%%%%%%%%%%%%%%%%%%
%%%%%%%%%%%%%%%%%%%%%%%%%%%%%%
%%%%%%%%%%%%%%%%%%%%%%%%%%%%%%%%
\subsection{Action of the Super-Poincar\'e algebra}
\label{Subsec:SC_Glob_Sym}
Let us now turn to the algebra of global symmetries and reformulate it in the homotopy algebra language, so that it is compatible with the two descriptions provided above.
We propose an interpretation of the super-Poincar\'e algebra in terms of a homotopy action on the $L_\infty$ algebra $(\cX_{\rm SYM},\{B_n\})$, which generalizes the familiar concept of a representation of a Lie algebra. 
Within this language, relations typically hold only up to homotopy terms. In the present context, these will encode well-known features about the action of symmetries in a field theory, such as global symmetries closing only on-shell, or up to gauge transformations.

All these `homotopy type' maps  may be gathered under the action of a newly defined set of multilinear operators
\begin{equation}
H_{m,n}(\Xi_1,\dots,\Xi_m)\,: \cX_{\rm SYM}^{\otimes n}\;\longrightarrow\;\cX_{\rm SYM}\;,   
\end{equation}
depending on $m$ parameters $\{\Xi_i\}_{i=1}^m$ of the super-Poincar\'e algebra and acting on $n$ elements of the $L_\infty$ algebra  $\cX_{\rm SYM}$. This feature of the homotopy algebra framework allows for the concept of \emph{representations up to homotopy} of the global symmetry algebra, whose significance and importance will become clearer at the end of the current subsection.

Finally, upon removing the color degrees of freedom and adding the $b$ operator, as discussed in the previous subsection, we will determine the super-Poincar\'e action on the kinematic algebra of super Yang-Mills theory.  This will allow us to determine the fate of the global (super)symmetries under the double copy.

\subsubsection*{Action on the $L_\infty$ algebra}

In the context of standard representation theory, a vector space $X$ carries a representation of a Lie algebra $\cG$ if, for any element $\Xi\in\cG$, there is a linear map $\Sigma(\Xi):X\rightarrow X$ obeying the usual closure relation $[\Sigma(\Xi_1),\Sigma(\Xi_2)]=\Sigma([\Xi_1,\Xi_2]_{\cG})$. In a field theory, however, the symmetry transformations $\delta_\Xi\varphi^i=\cR^i(\Xi,\varphi)$ can be nonlinear in the fields. This leads to a first generalization of the usual concept of representation.

Let us define a set of (possibly infinitely many) graded-symmetric $n$-linear maps
\begin{equation}\label{Def:SC_Sigmas}
    \Sigma_n(\Xi) \;:\; \cX_{\rm SYM}^{\otimes n} \longrightarrow \cX_{\rm SYM} \;,\quad |\Sigma_n(\Xi)| = 0 \;, \quad n\ge 1\;,
\end{equation}
depending on the global super-Poincar\'e parameter $\Xi=(\xi^\mu,\epsilon^\alpha)$ and acting on $n$ elements of the chain complex $\cX_{\rm SYM}$. Their action on fields $\cA$ can be defined by expanding the symmetry variations in powers of $\cA$ via the relation
\begin{equation}
    \delta_\Xi\cA \eqdef \Sigma_1(\Xi\vbar\cA) + \frac12 \Sigma_2(\Xi\vbar\cA,\cA) + \cdots \;.
\end{equation}
At lowest order, demanding that the field equations `rotate' covariantly under the global symmetries and that the latter are compatible with the gauge transformations, requires that the linear map $\Sigma_1(\Xi)$ commutes with the $L_\infty$ differential $B_1$.
By demanding that $[B_1, \Sigma_1(\Xi)]=0$ in general, one can extend the definition of $\Sigma_1(\Xi)$ to the whole chain complex $\cX_{\rm SYM}$, resulting in
\begin{subequations}\label{Eq:Diff_SC}\begin{align}
    \Sigma_1(\xi\,|\,\Uplambda) &= \bpm \xi\cdott\del\Lambda \\ \varnothing \epm \,\in\,X_{-1} \;, \\[2mm]
    %% %% %%
    \Sigma_1(\xi\,|\,\cA) &=
    \bpm \xi\cdott\del A_\mu + \del_\mu\xi\cdott A \;,\; \xi\cdott\del\chi + \frac14\omega_{\mu\nu}\gamma^{\mu\nu}\chi \\[1mm] \xi\cdott\del\varphi \epm \, \in\,X_0\;, \\[2mm]
    %% %% %%
    \Sigma_1(\xi\,|\,\cE) &=
    \bpm \xi\cdott\del E \;,\; \xi\cdott\del\cE_\chi + \frac14\omega_{\mu\nu}\gamma^{\mu\nu}\cE_\chi \\[1mm] \xi\cdott\del E_\mu + \del_\mu\xi^\nu E_\nu \;,\; \xi\cdott\del E_\chi + \frac14\omega_{\mu\nu}\gamma^{\mu\nu}E_\chi \epm \, \in\,X_1 \;, \\[2mm]
    %% %% %%
    \Sigma_1(\xi\,|\,\cN) &= \bpm \xi\cdott\del\cN_\chi + \frac14\omega_{\mu\nu}\gamma^{\mu\nu}\cN_\chi \\[1mm] \xi\cdott\del N \;,\; \xi\cdott\del N_\chi + \frac14\omega_{\mu\nu}\gamma^{\mu\nu}N_\chi \epm \,\in\,X_2\;, 
\end{align}\end{subequations}
when acting with a Poincar\'e transformation with parameter $\xi^\mu$, while it yields
\begin{subequations}\label{Eq:SUSY_SC_linear}\begin{gather}
    \Sigma_1(\epsilon\,|\,\Uplambda) = 0 \;,    \\[3mm]
    % % % %
    \Sigma_1(\epsilon\,|\,\cA_B) =
    \bpm \del_\mu A_\nu\gamma^{\mu\nu}\epsilon \\[1mm] \varnothing \epm \, \in\,X_0^{\rm F}\;, \quad
    \Sigma_1(\epsilon\,|\,\cA_F) = \bpm i\,\epsilon\gamma_\mu\chi \\[1mm] 0 \epm \,\in\,X_0^{\rm B} \\[2mm]
    % % % %
    \Sigma_1(\epsilon\,|\,\cE_B) =
    \bpm i\,(E_\mu-\del_\mu E)\gamma^\mu\epsilon \\[1mm] \del_\mu E_\nu\gamma^{\mu\nu}\epsilon \epm \, \in\,X_1^{\rm F}\;, \quad
    \Sigma_1(\epsilon\,|\,\cE_F) = \bpm \epsilon\,\cE_\chi \\[1mm] i\,\epsilon\gamma_\mu E_\chi \epm \,\in\,X_1^{\rm B}    \\[2mm]
    % % % %
    \Sigma_1(\epsilon\,|\,\cN_B) =
    \bpm N\epsilon \\[1mm] -i\,\delslash N\epsilon \epm \, \in\,X_2^{\rm F}\;, \quad
    \Sigma_1(\epsilon\,|\,\cN_F) = \bpm \varnothing \\[1mm] \epsilon\,N_\chi \epm \,\in\,X_2^{\rm B}    
\end{gather}\end{subequations}
when acting with a global supersymmetry with parameter $\epsilon^\alpha$. Let us point out that requiring $[B_1, \Sigma_1(\Xi)]=0$ does not fix $\Sigma_1(\Xi)$ completely, since one can shift it by an exact term, i.e.~$\Sigma_1(\Xi)\rightarrow\Sigma_1(\Xi)+[B_1,\Theta_1(\Xi)]$. The particular form \eqref{Eq:Diff_SC}, \eqref{Eq:SUSY_SC_linear} has been chosen so that the maps $\Sigma_1(\Xi)$ also commute with the $b$ operator, which will be relevant later on for the double copy.

Insisting that the theory is covariant under the action of the global symmetries, order by order in fields, gauge parameters, etc., leads one to postulate the following relations between the multilinear maps $\Sigma_n(\Xi)$ and the previously defined $L_\infty$ brackets $B_k$:
\begin{subequations}\begin{align}
    &B_1\Sigma_1(\Xi|x)=\Sigma_1\big(\Xi|B_1(x)\big)\;, \\[3mm]
    &B_1\Sigma_2(\Xi|x_1,x_2)+B_2\big(\Sigma_1(\Xi|x_1),x_2\big)+B_2\big(x_1,\Sigma_1(\Xi|x_2)\big) \nonumber \\
    &=\Sigma_1\big(\Xi|B_2(x_1,x_2)\big)+\Sigma_2\big(\Xi|B_1(x_1),x_2\big)+(-1)^{x_1}\Sigma_2\big(\Xi|x_1,B_1(x_2)\big)\;, \\[3mm]
    &B_1\Sigma_3(\Xi|x_1,x_2,x_3)+B_2\big(\Sigma_2(\Xi|x_1,x_2),x_3\big)+\text{two permutations of 123} \nonumber \\
    &+B_3\big(\Sigma_1(\Xi|x_1),x_2,x_3\big) + \text{two permutations of 123} \\
    &=\Sigma_1\big(\Xi|B_3(x_1,x_2,x_3)\big)+\Sigma_2\big(\Xi|B_2(x_1,x_2),x_3\big) + \text{two permutations of 123} \;, \nonumber \\[1mm]
    &\;\; \vdots \nonumber
\end{align}\end{subequations}
where $x_i\in\cX_{\rm SYM}$ are arbitrary elements of the $L_\infty$ algebra. These relations can be summarized as
\begin{equation}\label{BSigma compatibility}
    \sum_{i+j = n}[B_i,\Sigma_j(\Xi)] =0  \;,\quad \forall \, n \ge 2 \;,
\end{equation}
in a compact notation, where any $k$-linear graded-symmetric map, with $k<n$, acts in all possible inequivalent ways on the symmetrized tensor product of the inputs $x_1\wedge\dots\wedge x_n$. 
As we have previously mentioned, the above relations formalize the perturbative covariance of the gauge theory under the action of a global symmetry transformation. 

As usual, one can probe the algebra of global symmetries by taking the commutator of two successive transformations. In the simple case where $\Sigma(\Xi)$ furnishes a representation of the symmetry algebra $\cG$, successive transformations close according to
\begin{equation}
[\Sigma(\Xi_1),\Sigma(\Xi_2)]=\Sigma([\Xi_1,\Xi_2]_{\cG})\;,    
\end{equation}
where $[\Xi_1,\Xi_2]_{\cG}$ is the Lie bracket in $\cG$. In a gauge theory, such simple closure relations can be relaxed in various ways. Typically, it can happen that the commutator of two global symmetries closes only on-shell and up to gauge transformations. Given the set of maps $\Sigma_n(\Xi)$ and the $L_\infty$ brackets, we determine the first closure relations to be
\begin{subequations}\label{Eq:GlobSym_SC_Closure_Relations}\begin{align}
    [\Sigma_1(\Xi_1),\Sigma_1(\Xi_2)] &= \Sigma_1(\Xi_{12}) + [B_1,H_{2,1}(\Xi_1,\Xi_2)] \;, \\[2mm]
    2\,[\Sigma_1(\Xi_{[1}),\Sigma_2(\Xi_{2]})] &= \Sigma_{2}(\Xi_{12}) + [B_1,H_{2,2}(\Xi_1,\Xi_2)] + [H_{2,1}(\Xi_1,\Xi_2),B_2] \;,
\end{align}\end{subequations}
where $\Xi_{12} = [\Xi_1\,,\Xi_2]_{\rm s-Poinc}$ is the composite parameter obtained from the super-Poincar\'e bracket. We see that strict closure of the maps $\Sigma_n(\Xi_i)$ is relaxed by terms involving some homotopies $H_{m,n}$ of degree $|H_{m,n}|=1-m$. These take as input $m$ parameters $\{\Xi_i\}$ of the global symmetry algebra and $n$ elements of the $L_\infty$ algebra $\cX_{\rm SYM}$, and give back as output one element of $\cX_{\rm SYM}$:
\begin{equation}\label{Def:SC_H}
    H_{m,n}(\Xi_1, \dots, \Xi_m) \;:\; \cX_{\rm SYM}^{\otimes n} \longrightarrow \cX_{\rm SYM} \;, \quad m,n\ge 1 \;.
\end{equation}

Let us analyze closure of the supersymmetry algebra using the relations \eqref{Eq:GlobSym_SC_Closure_Relations}, which in the specific case  read
\begin{subequations}\label{Eq:SUSY_SC_Closure_Relations}\begin{align}
    &[\Sigma_1(\epsilon_1),\Sigma_1(\epsilon_2)] = \Sigma_1(a_{12}) + [B_1,H_{2,1}(\epsilon_1,\epsilon_2)] \;, \\[2mm]
    &[\Sigma_1(\epsilon_1),\Sigma_2(\epsilon_2)] \!-\! [\Sigma_1(\epsilon_2),\Sigma_2(\epsilon_1)] = [B_1,H_{2,2}(\epsilon_1,\epsilon_2)] + [H_{2,1}(\epsilon_1,\epsilon_2),B_2] \;,
\end{align}\end{subequations}
where $a^\mu_{12} \defeq -2i\,\epsilon_1\gamma^\mu\epsilon_2$ is the global translation parameter arising after performing two subsequent supersymmetry transformations. Upon defining the antisymmetric constant tensors
\begin{subequations}\begin{align}
    \epsilon_{12}^{\mu[n]} &\defeq \epsilon_1\gamma^{\mu[n]}\epsilon_2 \equiv \epsilon_1\gamma^{\mu_1\dots\mu_n}\epsilon_2\;, \\[2mm]
    \slashed{\Upsilon}_{12} &\defeq \frac78\, \epsilon_{12}^\mu\gamma_\mu - \frac1{16}\frac1{5!}\epsilon_{12}^{\mu[5]}\gamma_{\mu[5]} \;, 
\end{align}\end{subequations}
we can write down the linearized closure relations for all the elements of the chain complex \vspace{-5mm}
\begin{subequations}\label{Eq:SUSY_SC_linear_Closure}\begin{align}
    [\Sigma_1(\epsilon_1),\Sigma_1(\epsilon_2)]\Uplambda &= \Sigma_1(a_{12} \vbar \Uplambda) + H_{2,1}(\epsilon_1,\epsilon_2 \vbar B_1\Uplambda) \\[3mm]
    % % % %
    [\Sigma_1(\epsilon_1),\Sigma_1(\epsilon_2)]\cA_{\rm B} &= \Sigma_1(a_{12} \vbar \cA_{\rm B})) + B_1H_{2,1}(\epsilon_1,\epsilon_2 \vbar \cA_{\rm B}) + H_{2,1}(\epsilon_1,\epsilon_2 \vbar B_1\cA_{\rm B}) \\[3mm]
    % % % %
    [\Sigma_1(\epsilon_1),\Sigma_1(\epsilon_2)]\cA_{\rm F} &= \Sigma_1(a_{12} \vbar \cA_{\rm F})) + H_{2,1}(\epsilon_1,\epsilon_2 \vbar B_1\cA_{\rm F}) \\[3mm]
    % % % %
    [\Sigma_1(\epsilon_1),\Sigma_1(\epsilon_2)]\cE_{\rm B} &= \Sigma_1(a_{12} \vbar \cE_{\rm B})) + B_1H_{2,1}(\epsilon_1,\epsilon_2 \vbar \cE_{\rm B}) \\[3mm]
    % % % %
    [\Sigma_1(\epsilon_1),\Sigma_1(\epsilon_2)]\cE_{\rm F} &= \Sigma_1(a_{12} \vbar \cE_{\rm F})) + B_1H_{2,1}(\epsilon_1,\epsilon_2 \vbar \cE_{\rm F}) + H_{2,1}(\epsilon_1,\epsilon_2 \vbar B_1\cE_{\rm F}) \\[3mm]
    % % % %
    [\Sigma_1(\epsilon_1),\Sigma_1(\epsilon_2)]\cN_{\rm B} &= \Sigma_1(a_{12} \vbar \cN_{\rm B})) \\[3mm]
    % % % %
    [\Sigma_1(\epsilon_1),\Sigma_1(\epsilon_2)]\cN_{\rm F} &= \Sigma_1(a_{12} \vbar \cN_{\rm F})) + B_1H_{2,1}(\epsilon_1,\epsilon_2 \vbar \cN_{\rm F}) + H_{2,1}(\epsilon_1,\epsilon_2 \vbar B_1\cN_{\rm F}) \\[2mm]
    &\;\;\vdots \nonumber
\end{align}\end{subequations}
where the explicit action of the homotopy maps $H_{2,1}(\Xi_1,\Xi_2)$ is given by
\begin{subequations}\label{Eq:SUSY_SC_linear_Homotopies}\begin{align}
    \Sigma_1(a_{12} \vbar -) &\defeq a_{12}^\mu\del_\mu (-) \;, \\[3mm]
    %% On fields
    H_{2,1}(\epsilon_1,\epsilon_2 \vbar \cA_{\rm B}) &\defeq \bpm \Lambda_{12} \\[1mm] \varnothing \epm \in X_{-1}^{\rm B} \;, & \Lambda_{12}&\defeq 2i\,\epsilon_{12}^\mu\,A_\mu \;, \\[3mm]
    %% On bos e.o.m.'s
    H_{2,1}(\epsilon_1,\epsilon_2 \vbar \cE_{\rm B}) &\defeq \bpm 0 \\[1mm] \varphi_{12} \epm \in X_0^{\rm B} \;, & \varphi_{12}&\defeq -2i\,\epsilon_{12}^\mu\, E_\mu \;, \\[3mm]
    %% On fer e.o.m.'s
    H_{2,1}(\epsilon_1,\epsilon_2 \vbar \cE_{\rm F}) &\defeq \bpm \chi_{12} \\[1mm] \varnothing \epm \in X_0^{\rm F} \;, & \chi_{12} &\defeq \slashed{\Upsilon}_{12}\cE_\chi \;, \\[3mm]
    %% On bos Noether id.'s
    H_{2,1}(\epsilon_1,\epsilon_2 \vbar \cN_{\rm F}) &\defeq \bpm \cE_{\chi}^{12} \\[1mm] E_{\chi}^{12} \epm \in X_1^{\rm F} \;, & &\hspace{-7.5mm}\begin{matrix}\cE_{\chi}^{12} \defeq \epsilon_{12}^\mu\gamma_\mu\,\cN_\chi \\[1mm] \;E_{\chi}^{12} \defeq - \slashed{\Upsilon}_{12}N_\chi \end{matrix} \;, \\[3mm]
    %% On fer Noether id.'s
    H_{2,1}(\epsilon_1,\epsilon_2 \vbar \cR_{\rm F}) &\defeq \bpm \cN_{\chi}^{12} \\[1mm] N_{\chi}^{12} \epm \in X_2^{\rm F}\;, & &\hspace{-8mm}\begin{matrix} \hspace{-4mm}\cN_{\chi}^{12} \defeq \epsilon_{12}^\mu\gamma_\mu\cR_\chi \\[1mm] \;N_{\chi}^{12} \defeq - \epsilon_{12}^\mu\gamma_\mu R_\chi \end{matrix} \;.
\end{align}\end{subequations}
For the sake of brevity, we report the closure at the quadratic level only limited to gauge parameters and fields, where it takes the following form:
\begin{subequations}\label{Eq:SUSY_SC_quadratic_Closure}\begin{align}
    2\,[\Sigma_1(\epsilon_{[1}),\Sigma_2(\epsilon_{2]})](\Uplambda,\Uplambda) &= 0 \;, \\[3mm]
    % % % %
    2\,[\Sigma_1(\epsilon_{[1}),\Sigma_2(\epsilon_{2]})](\cA_{\rm B},\cA_{\rm B}) &= H_{2,1}(\epsilon_1,\epsilon_2\vbar B_2(A,A)) + 2 B_2(A, H_{2,1}(\epsilon_1,\epsilon_2 \vbar A)) \;, \\[3mm]
    % % % %
    2\,[\Sigma_1(\epsilon_{[1}),\Sigma_2(\epsilon_{2]})](\cA_{\rm B},\cA_{\rm F}) &= H_{2,1}(\epsilon_1,\epsilon_2 \vbar B_2(A,\chi)) + B_2(\chi, H_{2,1}(\epsilon_1,\epsilon_2 \vbar A)) \;, \\[3mm]
    % % % %
    2\,[\Sigma_1(\epsilon_{[1}),\Sigma_2(\epsilon_{2]})](\cA_{\rm F},\cA_{\rm F}) &= H_{2,1}(\epsilon_1,\epsilon_2 \vbar B_2(\chi,\chi)) \;,
\end{align}\end{subequations}
the relevant homotopy maps being given by the previous formulas \eqref{Eq:SUSY_SC_linear_Homotopies}.
Notice that all the bilinear maps $H_{2,2}(\epsilon_1,\epsilon_2)$ are vanishing for this specific gauge theory.

An analogous reasoning holds for the simpler instance of closure of the Poincar\'e subalgebra, in which case the only non-vanishing maps $\Sigma_{n}(\xi)$ are the linear ones, and one obtains strict closure, as expected:
\begin{equation}
    [\Sigma_1(\xi_{1}),\Sigma_1(\xi_{2})] = \Sigma_1(\xi_{12}) \;,\quad \xi_{12} = \cL_{\xi_1}\xi_2^\mu \;.
\end{equation}
Finally, the commutator of a supersymmetry and a Poincar\'e transformation also closes strictly:
\begin{equation}
    [\Sigma_1(\xi),\Sigma_1(\epsilon)]= \Sigma_1(\epsilon_{\rm eff}) \;,
\end{equation}
in terms of a rotated global supersymmetry parameter
\begin{equation}
    \epsilon_{\rm eff} \defeq \frac14\omega_{\mu\nu}\gamma^{\mu\nu}\epsilon \;.
\end{equation}

\subsubsection*{Action on the kinematic algebra}

Having  the super-Poincar\'e action on the $L_\infty$ algebra $\cX_{\rm SYM}$, we will now strip off the color  and define a suitable action of global symmetries on the kinematic algebra $\cK_{\rm SYM}$. This can be achieved by defining  maps $\rho_n(\Xi)$ and homotopies $h_{m,n}(\Xi_1,\dots,\Xi_m)$ analogous to the $\Sigma_n(\Xi)$ and $H_{m,n}(\Xi_1,\dots,\Xi_m)$, acting on the kinematic algebra:
\begin{subequations}\begin{align}
    \rho_{n}(\Xi) \;&:\; \cK_{\rm SYM}^{\otimes n} \longrightarrow \cK_{\rm SYM} \;,\quad |\rho_{n}(\Xi)| = 1-n \;, \\[2mm]
    h_{m,n}(\Xi_1, \dots, \Xi_m) \;&:\; \cK_{\rm SYM}^{\otimes n} \longrightarrow \cK_{\rm SYM} \;,\quad |h_{m,n}(\Xi_1, \dots, \Xi_m)| = 2-m-n \;,
\end{align}\end{subequations}
with $m,n \ge 1$. Following the prescription implemented in \eqref{Eq:Linf_Cinf_relations}, the relation between the `representation maps' $\Sigma_n$ and $\rho_n$ and between homotopies $H_{m,n}$ and $h_{m,n}$ is pretty much the same as the relation between the $L_\infty$ brackets $B_n$ and $C_\infty$ products $m_n$: 
\begin{subequations}\label{Eq:Linf_Cinf_SUSY_relations}\begin{align}
    \Sigma_1(\Xi \vbar x) &= \rho_1(\Xi \vbar u^a)\otimes t_a \;, \\[2mm]
    \Sigma_2(\Xi \vbar x_1,x_2) &= (-1)^{x_1}f^a{}_{bc}\,\rho_2(\Xi \vbar u_1^b,u_2^c)\otimes t_a \;,
\end{align}\end{subequations}
where we expanded a generic element $x\in\cX_{\rm SYM}$ as $x=u^a\otimes t_a$, with $u^a\in\cK_{\rm SYM}$.
Analogously, the homotopy maps $h_{2,n}$ will be defined through
\begin{subequations}\label{Eq:Linf_Cinf_HOM_relations}\begin{align}
    H_{2,1}(\Xi_1,\Xi_2 \vbar x) &= h_{2,1}(\Xi_1,\Xi_2 \vbar u^a)\otimes t_a \;, \\[2mm]
    H_{2,2}(\Xi_1,\Xi_2 \vbar x_1,x_2) &= (-1)^{x_1}f^a{}_{bc}\,h_{2,2}(\Xi_1,\Xi_2 \vbar u_1^b,u_2^c)\otimes t_a \;.
\end{align}\end{subequations}

The maps and homotopies so defined implement a consistent action of global symmetries on the $C_\infty$ algebra on $\cK_{\rm SYM}$.
To access the relevant kinematic algebra, however, the first extra step is to introduce the $b$ operator, which we have discussed in detail in sec.~\ref{Subsec:SC_Hom_Form}.
The linear maps $\rho_1(\Xi)$ and $h_{2,1}(\Xi_1,\Xi_2)$ defined above are not a priori guaranteed to obey simple relations with $b$. For instance, from\footnote{This is the same relation as $[B_1,\Sigma_1(\Xi)]=0$ upon color-stripping.} $[m_1,\rho_1(\Xi)]=0$ one can only deduce that $[b,\rho_1(\Xi)]$ is $m_1$-closed, but this  in turn suggests that it is exact, 
$[b,\rho_1(\Xi)]=[m_1,\Theta_1(\Xi)]$, with $\Theta_1(\Xi)$  a further homotopy of degree $-2$, as shown in \cite{Bonezzi:2025anl} for the case of $\cN=4$ super-Yang-Mills theory in four dimensions.
Both $\rho_1(\Xi)$ and $h_{2,1}(\Xi_1,\Xi_2)$ are defined as equivalence classes, since they can be shifted by $m_1$-exact terms while preserving their defining properties. It turns out that we can pick representatives in which the commutation relations hold strictly, namely
\begin{equation}\label{Cinf_Comm_Conditions}
    [b,\rho_1(\Xi)] = [b,h_{2,1}(\Xi_1,\Xi_2)] = 0\;,
\end{equation}
by defining the action of the super-Poincaré maps as in \eqref{Eq:SUSY_SC_linear}.
This feature will turn out to be essential for the construction of the double copy, where a constraint involving the operator $b$ will be imposed in order to obtain the correct spectrum of fields in the final theory.
We want to emphasize again that in the homotopy algebra language relations like \eqref{Cinf_Comm_Conditions} rarely hold strictly, but typically  only up to homotopy. This characteristic allows for a class of \emph{homotopy equivalent representations} of a given map and permits a certain freedom of choice between them, i.e.~we can pick the most convenient representation.

%%%%%%%%%%%%%%%%%%%%%%%%%%%%%%%%%%
%%%%%%%%%%%%%%%%%%%%%%%%%%%%%%%%
%% BEGIN SECTION 4 %%%%%%%%%%%%%%%
%%%%%%%%%%%%%%%%%%%%%%%%%%%%%%%%
%%%%%%%%%%%%%%%%%%%%%%%%%%%%%%%%%%
\section{Maximal Supergravity in $D=10$}
\label{Sec:DC_Sugra}
In this  section we perform the double copy of 10-dimensional super Yang-Mills theory in order to obtain a \emph{super double field theory} (SDFT) version of ${\cal N} = 2$ supergravity (see \cite{Hohm:2011zr,Hohm:2011dv,Jeon:2011sq,Jeon:2012hp} for type II double field theories). We analyze the resulting theory from the point of view of both the local dynamics and the action of global symmetries. The resultant theory contains the gauge symmetries corresponding to linearized diffeomorphisms and local supersymmetries, as well as $p$-form gauge symmetries. We show that the double copy also exhibits a doubled super-Poincaré algebra of global symmetries, which is expected from the perturbative expansion of a double field  theory around flat space.
Subsequently, we will show that with a convenient identification of the spin groups as well as with a proper redefinition of the double-copied fields, we retrieve both type IIA and type IIB supergravities.

%%%%%%%%%%%%%%%%%%%%%%%%%%%%%%%%
%%%%%%%%%%%%%%%%%%%%%%%%%%%%%%
%%%%%%%%%%%%%%%%%%%%%%%%%%%%%%%%
\subsection{Double Copy and Local Symmetries: The Spectrum}
Let us begin by recalling that our prescription for the double copy requires taking the tensor product of two copies $\cK_{\rm SYM}$ and $\widetilde\cK_{\rm SYM}$ of the super Yang-Mills kinematic algebra \cite{Bonezzi:2022yuh}. At a practical level, this means taking the tensor products of all the elements of the two copies of the $C_\infty$ chain complex and rearranging them in a bigger chain complex, which pertains to a new $L_\infty$ algebra upon imposing the \emph{section constraint}
\begin{equation}\label{SecConstr}
    \Box \equiv \widetilde\Box \;,
\end{equation}
also known as \emph{strong constraint} or \emph{level-matching condition} in closed-string theory \cite{Hull:2009mi}. This can be solved by identifying the two sets of coordinates $x^\mu$ and $\tilde x^\tmu$ of the two copies of the gauge theory kinematic algebra.
The process can be schematized as follows:
\begin{equation*}
    \left.\begin{array}{cc}
        \cX_{\rm SYM} = \cK_{\rm SYM}\otimes\mathfrak{g} \;\xrightarrow{\text{color-strip}}\; \cK_{\rm SYM} \; \\
        \\
        \widetilde\cX_{\mathrm{SYM}} = \widetilde\cK_{\rm SYM}\otimes\widetilde{\mathfrak{g}} \;\xrightarrow{\text{color-strip}}\; \widetilde\cK_{\rm SYM} \;
    \end{array}\right\}
    \;\xrightarrow[\text{product}]{\text{tensor}}\; 
    \cX_{\mathrm{SDFT}}
    \defeq \big(\cK_{\rm SYM}\otimes\widetilde{\cK}_{\rm SYM}\big)\Big\rvert_{\rm section}\;.
\end{equation*}
Lastly, we need to impose a further constraint so that the final complex contains the correct number of elements and matches the standard DFT complex for ${\cal N} = 2$ supergravity. To this end, we require all elements $\Omega\in\cK_{\rm SYM}\otimes\widetilde{\cK}_{\rm SYM}$ of the SDFT chain complex to obey $b^-\Omega = 0$, having defined a new nilpotent operator
\begin{equation}
    b^-\defeq \tfrac12\,\big(b\otimes\tilde\1-\1\otimes\tilde b\big)\;, \quad (b^-)^2=0\;.
\end{equation}
This new operator induces a degree shift on the elements of the double-copy chain complex, in analogy with the way $b$ behaves in the single-copy (gauge theory) case.

Applying this procedure to 10-dimensional super Yang-Mills, one retrieves a bigger -- still semi-infinite -- chain complex for a SDFT version of $\cN = 2$ supergravity related to a new $L_\infty$ algebra $(\cX_{\rm SDFT}, \bbB_n)$:
\begin{equation}
\label{CC:SuGra}
   \begin{tikzcd}[row sep=0pt, column sep={2.1cm,between origins}]
        X_{-2} \arrow[r,"\bbB_1"] &X_{-1} \arrow[r,"\bbB_1"] & X_{0} \arrow[r,"\bbB_1"]  &X_{1}\arrow[r,"\bbB_1"] &X_{2}\arrow[r,"\bbB_1"] &X_{3} \arrow[r,"\bbB_1"] & \cdots \\
        %%%
        \mathbbg{t} & \bbLambda & \mathbb{H} & \bbE & \bbN  & \mathbb{R} & \cdots
   \end{tikzcd}\quad,
\end{equation}
where $\mathbbg{t}$, $\bbLambda$, $\mathbb{H}$, $\bbE$, $\bbN$ and $\mathbb{R}$ respectively are the sets of gauge-for-gauge parameters, gauge parameters, fields, equations of motion, Noether identities and Noether-for-Noether identities of the resulting theory. We keep the convention in which the fields stay in the subspace $X_0$ of $L_\infty$ degree zero.

The new relevant double-copy $L_\infty$ $n$-brackets $\bbB_n$ are defined as \cite{Bonezzi:2022bse}
\begin{subequations}\label{Def:DC_B}\begin{align}
    \bbB_1 &\defeq m_1\otimes\tilde\1 + \1\otimes\tilde m_1 \label{Def:DC_B1}\;, \\[1mm]
    \bbB_2 &\defeq -\tfrac12 \,b^-(m_2\otimes\tilde m_2)\;, \label{Def:DC_B2} \\
    &\;\;\vdots \nonumber
\end{align}\end{subequations}
with respect to the kinematic algebra $n$-linear maps $m_n$ and allow us to describe the double-copy theory in terms of field equations expressed in the Maurer-Cartan form
\begin{equation}
    \bbB_1(\bbH) + \frac12\,\bbB_2(\bbH,\bbH) + \frac16\,\bbB_3(\bbH,\bbH,\bbH) + \cdots = 0 \;,
\end{equation}
which are covariant under the gauge transformations
\begin{equation}
    \delta_\bbLambda\bbH = \bbB_1(\bbLambda) + \bbB_2(\bbLambda,\bbH) + \cdots \;.
\end{equation}
Let us now do a more detailed analysis of the super DFT spectrum that we have obtained.

\subsubsection*{Gauge Parameters}
The set of gauge parameters
\begin{equation}\label{DC_Gauge_Param}
    \bbLambda = \bpm \Lambda_{\mu} \,,\tilde\Lambda_{\tilde\mu} \\[1mm] \eta\epm
    \oplus
    \bpm \varepsilon^\talpha \,, \tilde\varepsilon^\alpha \\[1mm] \varnothing \epm
\end{equation}
consists of a bosonic and a fermionic sector. The bosonic sector contains two vectors and one scalar, which admit trivial parameters of the form
\begin{equation}
    \Lambda_\mu^{\rm triv} = \del_\mu\tau\;, \quad
    \tilde\Lambda_{\tilde\mu}^{\rm triv} = \tilde\del_{\tilde\mu}\tau\;, \quad
    \eta^{\rm triv} = \Box\tau \;.
\end{equation}
The fermionic sector, instead, consists of the two local supersymmetry parameters $\varepsilon$, $\tilde\varepsilon$.

\subsubsection*{Field Content}
The field content can be studied in its usual three different sectors, that are Neveu–Schwarz--Neveu–Schwarz (NS-NS), Neveu–Schwarz--Ramond (NS-R) and Ramond--Ramond (R-R):
\begin{equation}\label{DC_Field_NSNS}
    \bbH = \bbH_{\rm NS-NS} \oplus \bbH_{\rm NS-R} \oplus \bbH_{\rm R-R} \;.
\end{equation}
The NS-NS sector
\begin{equation}
    \bbH_{\rm NS-NS} = \bpm e_{\mu\tnu} \,, e \,, \tilde e \\[1mm]
    f_\mu \,, \tilde f_{\tilde\mu} \epm
\end{equation}
consists of a tensor $e_{\mu\tnu}$ and two scalars $e$, $\tilde e$, which give rise to the graviton, B-field and dilaton after identifying the spacetime coordinates. It also includes two auxiliary vector fields $f_\mu$, $\tilde f_\tmu$, similar to the original formulation of DFT from string field theory \cite{Hull:2009mi}. Altogether, they obey the linearized field equations $\bbB_1(\bbH_{\rm NS-NS}) = 0$ given by
\begin{subequations}\label{DCEomNSNS}\begin{align}
    \Box e_{\mu\tnu} + \tilde\del_{\tilde\nu}f_\mu - \del_\mu\tilde f_{\tilde\nu} &= 0\;, & 
    & \\[1mm]
    \del_\mu\tilde e - \tilde\del^{\tilde\nu}e_{\mu\tilde\nu} - f_\mu &= 0\;, & 
    \tilde\del_{\tilde\nu}e + \del^\mu e_{\mu\tilde\nu} - \tilde f_{\tilde\nu} &= 0\;, \\[1mm]
    \Box e - \del^\mu f_\mu &= 0\;, & 
    \Box\tilde e - \tilde\del^{\tilde\mu}\tilde f_{\tilde\mu} &= 0\;,
\end{align}\end{subequations}
which are invariant under the linearized gauge transformations $\bbB_1(\bbLambda) = \delta_\bbLambda^{\rm lin}\bbH_{\rm NS-NS}$:
\begin{subequations}\label{DCgaugeNSNS}\begin{align}
    \delta_\bbLambda e_{\mu\tilde\nu} &= \del_\mu\tilde\Lambda_{\tilde\nu} - \tilde\del_{\tilde\nu}\Lambda_\mu\;,\ &
    & \\[1mm]
    \delta_\bbLambda e &= \del^\mu\Lambda_\mu - \eta\;, &
    \delta_\bbLambda\tilde e &= \tilde\del^{\tilde\mu}\tilde\Lambda_{\tilde\mu} - \eta\;, \\[1mm]
    \delta_\bbLambda f_\mu &= \Box\Lambda_\mu - \del_\mu\eta\;, &
    \delta_\bbLambda\tilde f_{\tilde\mu} &= \Box\tilde\Lambda_{\tilde\mu} - \tilde\del_{\tilde\mu}\eta\;.
\end{align}\end{subequations}
The fermions belong to the NS-R sector, whose fields
\begin{equation}
    \bbH_{\rm NS-R} = \bpm \psi_\mu{}^\talpha, \rho_\talpha \,; \, \tilde \psi_\tmu{}^\alpha, \tilde\rho_\alpha \\[1mm]
    \varphi^\talpha \,; \,\tilde\varphi^\alpha \epm
\end{equation}
comprise the two gravitini $\psi_\mu{}^\talpha$, $\psi_\tmu{}^\alpha$ and the two DFT dilatini $\rho_\talpha$, $\tilde\rho_\alpha$ together with a couple of auxiliary fermions $\varphi^\talpha$, $\tilde\varphi^\alpha$. Their linearized equations of motion read
\begin{subequations}\label{DC_Eom_NSR}\begin{align}
    \gamma^\tnu\tilde\del_\tnu\psi_\mu - \del_\mu\rho &= 0\;, &
    \gamma^\nu\del_\nu\tilde\psi_{\tilde\mu} - \tilde\del_{\tilde\mu}\tilde{\rho} &= 0\;, \\[1mm]
    \del^\mu\psi_\mu - \varphi &= 0\;, &
    \tilde\del^{\tilde\mu}\tilde\psi_{\tilde\mu} - \tilde\varphi &= 0\;, \\[1mm]
    \gamma^\tmu\tilde\del_\tmu\rho - \varphi &= 0\;, &
    \gamma^\mu\del_\mu\tilde\rho - \tilde\varphi &= 0\;,
\end{align}\end{subequations}
and are invariant under the linearized transformations
\begin{subequations}\label{DC_gauge_NSR}\begin{align}
    \delta_\bbLambda\psi_\mu &= \del_\mu\varepsilon\;, &
    \delta_\bbLambda\tilde\psi_{\tilde\mu} &= \tilde\del_{\tilde\mu}\tilde\varepsilon\;, \\[1mm]
    \delta_\bbLambda\rho &= \gamma^\tmu\tilde\del_\tmu\varepsilon\;, &
    \delta_\bbLambda\tilde\rho &=\gamma^\mu\del_\mu\tilde\varepsilon\;, \\[1mm]
    \delta_\bbLambda\varphi &= \Box\varepsilon\;, &
    \delta_\bbLambda\tilde\varphi &= \Box\tilde\varepsilon\;.
\end{align}\end{subequations}
Finally, the R-R sector 
\begin{equation}
    \bbH_{\rm R-R} = \bpm F^{\alpha\tbeta} \\[1mm]
    \varnothing \epm
\end{equation}
contains only a gauge-invariant bispinor obeying the Bargmann–Wigner equations, i.e. the massless Dirac equation in both indices
\begin{equation}\label{DC_Eom_RR}
    \gamma^\tmu\tilde\del_\tmu F = 0\;, \quad\quad \gamma^\mu\del_\mu F = 0\;.
\end{equation}
In the basis described above, all fields and gauge parameters are real.

%%%%%%%%%%%%%%%%%%%%%%%%%%%%%%%%
%%%%%%%%%%%%%%%%%%%%%%%%%%%%%%
%%%%%%%%%%%%%%%%%%%%%%%%%%%%%%%%
\subsection{Double Copy of Global Symmetries}
Let us now analyze how global symmetries behave after the double copy. We can define a new collective global parameter
\begin{equation}
    \bfXi \defeq (\Xi,\tilde\Xi) \equiv (\upxi, \upepsilon) \;,
\end{equation}
where $\upxi^M \defeq (\xi^\mu,\tilde\xi^\tmu)$ and $\upepsilon \defeq (\epsilon,\tilde\epsilon)$ respectively are the global parameters for doubled Poincar\'e and global supersymmetry transformations. In the same fashion as for \eqref{Def:DC_B}, we can define the new $n$-linear maps
\begin{subequations}\begin{align}
    \bbSigma_1(\bfXi) &\defeq \rho_1(\Xi)\otimes\tilde{\1} + \1\otimes\tilde{\rho}_1(\tilde\Xi)\;, \\[1mm]
    \bbSigma_2(\bfXi) &\defeq \tfrac12\,b^-\big(\rho_2(\Xi)\otimes\tilde m_2+m_2\otimes\tilde\rho_2(\tilde\Xi)\big)\;,
\end{align}\end{subequations}
which determine the action of the doubled global (super)symmetries on the super-DFT fields. By construction, they obey the covariance relations
\begin{equation}\label{double global covariance}
[\bbB_1,\bbSigma_1(\bfXi)]=0 \;,\quad [\bbB_1,\bbSigma_2(\bfXi)]=[\bbSigma_1(\bfXi),\bbB_2] \;,
\end{equation}
analogous to the gauge theory ones \eqref{BSigma compatibility}.
They also obey the homotopy closure relations \hspace{-5mm}
\begin{subequations}\begin{align}
[\bbSigma_1(\bfXi_1),\bbSigma_1(\bfXi_2)] &= \bbSigma_1(\bfXi_{12}) + [\bbB_1, \bbH_{2,1}(\bfXi_1,\bfXi_2)] \\[1mm]
    2\,[\bbSigma_1(\bfXi_{[1}), \bbSigma_2(\bfXi_{2]})] &= \bbSigma_2(\bfXi_{12}) + [\bbB_1,\, \bbH_{2,2}(\bfXi_1,\bfXi_2)] + [\bbH_{2,1}(\bfXi_1,\bfXi_2),\, \bbB_2]
\end{align}\end{subequations}
where $\bbH_{m,n}(\bfXi_1,\dots,\bfXi_m)$ are \emph{homotopy $n$-linear maps} playing in the double-copy theory the same role as their single-copy counterpart \eqref{Def:SC_H}. They can also be constructed from the gauge theory kinematic products and super-Poincar\'e action as follows 
\begin{subequations}\begin{align}
    \bbH_{2,1}(\bfXi_1,\bfXi_2) &\defeq h_{2,1}(\Xi_1,\Xi_2) \otimes\tilde\1 + \1\otimes\tilde{h}_{2,1}(\tilde\Xi_1,\tilde\Xi_2) \;, \label{Eq:DC_Glob_Sym_Rel_1}\\[1mm]
    \bbH_{2,2}(\bfXi_1,\bfXi_2) &\defeq -\tfrac12\,b^- \Big( h_{2,2}(\Xi_1,\Xi_2) \otimes \tilde m_2 + m_2 \otimes \tilde h_{2,2}(\tilde\Xi_1,\tilde\Xi_2) \label{Eq:DC_Glob_Sym_Rel_2}\\[1mm]
    &\hspace{1.8cm}+ \rho_2(\Xi_1)\otimes\tilde\rho_2(\tilde\Xi_2) - \rho_2(\Xi_2)\otimes\tilde\rho_2(\tilde\Xi_1) \Big) \;.\nonumber
\end{align}\end{subequations}
Let us now show that the first of the two closure relations holds:
\begin{equation}\begin{split}
    [\bbSigma_1(\bfXi_1),\bbSigma_1(\bfXi_2)] &= [\rho_1(\Xi_1)\otimes\tilde{\1} + \1\otimes\tilde{\rho}_1(\tilde\Xi_1),\, \rho_1(\Xi_2)\otimes\tilde{\1} + \1\otimes\tilde{\rho}_1(\tilde\Xi_2)] \\
    &= \bigl\{ \rho_1(\Xi_{12}) + [m_1,\, h_{2,1}(\Xi_1,\Xi_2)] \bigr\} \otimes \tilde{\1} \\
    &\quad + \1\otimes \bigl\{ \tilde\rho_1(\tilde\Xi_{12}) + [\tilde{m}_1,\, \tilde{h}_{2,1}(\tilde\Xi_1,\tilde\Xi_2)] \bigr\} \\
    &= \bigl\{ \rho_1(\Xi_{12})\otimes\tilde{\1} + [m_1\otimes\tilde{\1} + \1\otimes\tilde m_1,\, h_{2,1}(\Xi_1,\Xi_2)\otimes\tilde{\1}] \bigr\} \\
    &\quad + \bigl\{ \1\otimes\tilde\rho_1(\tilde\Xi_{12}) + [m_1\otimes\tilde\1 + \1\otimes\tilde{m}_1,\, \1\otimes\tilde{h}_{2,1}(\tilde\Xi_1,\tilde\Xi_2)] \bigr\} \\
    &= \bbSigma_1(\bfXi_{12}) + [\bbB_1, \bbH_{2,1}(\bfXi_1,\bfXi_2)] \;,
\end{split}\end{equation}
where in the third step we have used  that $\tilde m_1$ commutes with $h_{2,1}(\Xi_1,\Xi_2)$ and that $m_1$ commutes with $\tilde h_{2,1}(\tilde\Xi_1,\tilde\Xi_2)$.
We can also prove the second closure relation
\begin{equation}\label{Eq:DC_Glob_Sym_Rel_2_Proof}\begin{split}
    [\bbSigma_1&(\bfXi_1), \bbSigma_2(\bfXi_2)] - (1\leftrightarrow2) = \\
    &= \big[\rho_1(\Xi_1)\otimes\tilde{\1} + \1\otimes\tilde{\rho}_1(\tilde\Xi_1),\, \tfrac12\,b^-\big(\rho_2(\Xi_2)\otimes\tilde m_2 + m_2\otimes\tilde\rho_2(\tilde\Xi_2)\big)\big] - (1\leftrightarrow2) \\
    &= \tfrac12\,b^- \Big\{\Big( \rho_2(\Xi_{12}) + [m_1,\, h_{2,2}(\Xi_1,\Xi_2)] + [h_{2,1}(\Xi_1,\Xi_2),\,m_2]\Big) \otimes \tilde m_2 \\
    &\hspace{1.2cm} - \rho_2(\Xi_1) \otimes [\tilde m_1,\, \tilde\rho_2(\tilde\Xi_2)] - (1\leftrightarrow2) + [m_1,\,\rho_2(\Xi_1)] \otimes \tilde\rho_2(\tilde\Xi_2) - (1\leftrightarrow2) \\
    &\hspace{1.2cm} + m_2 \otimes \Big( \tilde\rho_2(\tilde\Xi_{12}) + [\tilde m_1,\, \tilde h_{2,2}(\tilde\Xi_1,\tilde\Xi_2)] + [\tilde h_{2,1}(\tilde\Xi_1,\tilde\Xi_2),\,\tilde m_2]\Big) \Big\} \\
    &= \bbSigma_2(\bfXi_{12}) + [\bbB_1,\, \bbH_{2,2}(\bfXi_1,\bfXi_2)] + [\bbH_{2,1}(\bfXi_1,\bfXi_2),\, \bbB_2]
\end{split}\end{equation}
using the same manipulations.
Notice that the proof \eqref{Eq:DC_Glob_Sym_Rel_2_Proof} uses the property
\begin{equation}\label{strict commutation}
    [b,\, \rho_1(\Xi)] = [b,\, h_{2,1}(\Xi_1,\Xi_2)] = 0 \;,
\end{equation}
which is the same condition as \eqref{Cinf_Comm_Conditions}. 
If \eqref{strict commutation} held only up to homotopy, one could still construct double copy maps $\bbSigma_n(\bfXi)$ obeying \eqref{double global covariance}. These, however, would not be compatible with the constraint $b^-\Omega=0$ imposed on the super-DFT chain complex. This, ultimately, was the reason for choosing the representatives obeying the strict condition \eqref{Cinf_Comm_Conditions}, as discussed at the end of Section \ref{Sec:10DKinAlg} and already pointed out in \cite{Bonezzi:2025anl}.

Finally, let us provide some explicit examples of global symmetry transformations for the free theory. Under doubled Poincar\'e, the transformations take the usual form
\begin{equation}
    \delta_\upxi\bbH = \cL_\upxi\bbH \;,
\end{equation}
$\cL_\upxi$ being the Lie derivative with respect to the parameter $\upxi^M$. More interestingly, instead, the linearized double copy theory shows invariance under the following global supersymmetry transformations:
\begin{fleqn}\begin{subequations}\label{DC_Glob_SUSY_trans}\begin{align}
    \delta_\upepsilon e_{\mu\tnu} &= i\,\tepsilon^\talpha\gamma_{\tnu\;\talpha\tbeta}\,\psi_\mu{}^\tbeta - i\,\epsilon^\alpha\gamma_{\mu\;\alpha\beta}\,\tilde\psi_\tnu{}^\beta \;, \hspace{-0.4cm}& & \\[1mm]
    \delta_\upepsilon e &= i\,\epsilon^\alpha\trho_\alpha \;, \hspace{-0.4cm}&
    \delta_\upepsilon {\tilde e}  &=i\, \tepsilon^\talpha\rho_\talpha \;, \\[1mm]
    \delta_\upepsilon f_\mu &= i\,\epsilon^\alpha\gamma_{\mu\;\alpha\beta}\,\tilde\varphi^\beta \;, \hspace{-0.4cm}& 
    \delta_\upepsilon \tilde f_\tmu &= i\,\tepsilon^\talpha\gamma_{\tmu\;\talpha\tbeta}\,\varphi^\tbeta \;; \\[2mm]
    % % % %
    \delta_\upepsilon \psi_\mu{}^\talpha &= \tdel_\tnu e_{\mu\trho}(\gamma^{\tnu\trho}\,\tepsilon)^\talpha -\epsilon^\beta\gamma_{\mu\;\beta\lambda}F^{\lambda\talpha} \;, \hspace{-0.4cm}& 
    \delta_\upepsilon \tpsi_\tmu{}^\alpha &= \!-\, \del_\nu e_{\rho\tmu}(\gamma^{\nu\rho}\,\epsilon)^\alpha \!-\tepsilon^\tbeta\gamma_{\tmu\;\tbeta\tlambda}F^{\alpha\tlambda} ,\!\!\!\! \\[1mm]
    \delta_\upepsilon \rho_\talpha &= \tilde f_\tmu\gamma^\tmu{}_{\talpha\tbeta}\,\tepsilon^\tbeta - \tdel_\tmu \te\gamma^\tmu{}_{\talpha\tbeta}\,\tepsilon^\tbeta \;, \hspace{-0.4cm}& 
    \delta_\upepsilon \trho_\alpha &= f_\mu\gamma^\mu{}_{\alpha\beta}\,\epsilon^\beta - \del_\mu e\gamma^\mu{}_{\alpha\beta}\,\epsilon^\beta \;, \\[1mm]
    \delta_\upepsilon \varphi^\talpha &= \tdel_\tmu\tilde f_\tnu\gamma^{\tmu\tnu\;\talpha}{}_\tbeta\,\tepsilon^\tbeta \;, \hspace{-0.4cm}& 
    \delta_\upepsilon \tilde\varphi^\alpha &= \del_\mu f_\nu\gamma^{\mu\nu\;\alpha}{}_\beta\,\epsilon^\beta \;; \\[2mm]
    % % % %
    \delta_\upepsilon F^{\alpha\tbeta} &= i\,\del_\mu\psi_\nu{}^\tbeta\gamma^{\mu\nu\;\alpha}{}_\lambda\epsilon^\lambda +i\, \tilde\del_\tmu\tilde\psi_\tnu{}^\alpha\gamma^{\tmu\tnu\;\tbeta}{}_\tlambda\tilde\epsilon^\tlambda \;, \hspace{-0.4cm}& &
\end{align}\end{subequations}\end{fleqn}
where the supersymmetry parameters $(\epsilon^\alpha,\tilde\epsilon^{\tilde\alpha})$ are real Majorana-Weyl spinors.
In the next subsection we will relate this theory to 10D supergravity.

%%%%%%%%%%%%%%%%%%%%%%%%%%%%%%%%
%%%%%%%%%%%%%%%%%%%%%%%%%%%%%%
%%%%%%%%%%%%%%%%%%%%%%%%%%%%%%%%
\subsection{Type IIA and Type IIB Supergravities}
We will now relate  type IIA and type IIB supergravities to the super double field theory that we have obtained through the double copy procedure. To this end, we  impose the section constraint \eqref{SecConstr} and identify the two sets of coordinates $x^\mu \equiv \tilde x^\tmu$ and derivatives $\del_\mu \equiv \tdel_\tmu$ as well as the respective Lorentz groups ${\rm SO}(1,9)\equiv\widetilde{\rm SO}(1,9)$. %Thereafter, we want to adopt the string frame by performing a few field redefinitions. 
Finally, the choice of the spinor representations and a detailed analysis of the field content will lead us to the two desired theories. 

\subsubsection*{Field Redefinitions}
Let us begin by looking at the NS-NS sector in order to retrieve the graviton $h_{\mu\nu}$, the $B$-field $B_{\mu\nu}$ and a gauge-invariant dilaton $\phi$. To this end, we can split the tensor $e_{\mu\nu}$ in its symmetric and antisymmetric parts
\begin{equation}
    e_{\mu\nu} = h_{\mu\nu} + B_{\mu\nu} \;,\quad\text{where}\quad h_{\mu\nu} \defeq e_{(\mu\nu)} \;,\quad B_{\mu\nu} \defeq e_{[\mu\nu]}\;.
\end{equation}
The graviton and the $B$-field linearized gauge transformations now read
\begin{equation}
    \delta_\bbLambda h_{\mu\nu} = \del_\mu\lambda_\nu + \del_\nu\lambda_\mu \;,\quad \delta_\bbLambda B_{\mu\nu} = \del_\mu\zeta_\nu - \del_\nu\zeta_\mu 
\end{equation}
having redefined the two vector gauge parameters $\Lambda_\mu$, $\tilde\Lambda_\mu$ as
\begin{equation}
    \lambda_\mu \defeq \tfrac12(\tilde\Lambda_\mu - \Lambda_\mu) \;,\quad \zeta_\mu \defeq \tfrac12(\tilde\Lambda_\mu + \Lambda_\mu)\;.
\end{equation}
The gauge invariant dilaton can be defined as a linear combination of the two scalar fields $e$, $\tilde e$ with the trace $h$ of the graviton $h_{\mu\nu}$
\begin{equation}
    \phi \defeq \tfrac12(\tilde e - e - h)\;.
\end{equation}
A couple of real gauge-invariant dilatini $\lambda_\talpha$, $\tlambda_\alpha$ is encoded in the linear combinations
\begin{equation}
    \lambda_\talpha \defeq \rho_\talpha - (\gamma^\mu)_{\talpha\tbeta}\,\psi_\mu{}^\tbeta \;,\qquad
    \tlambda_\alpha \defeq \trho_\alpha - (\gamma^\mu)_{\alpha\beta}\,\tilde\psi_\mu{}^\beta \;,
\end{equation}
where we have explicitly written all the spinor indices for the sake of clarity. In particular, notice that the identification of the two Lorentz groups does not identify $(\gamma^\mu)_{\alpha\beta}$ with $(\gamma^\mu)_{\talpha\tbeta}$, since we have not yet specified the spinor representations for the solution of the section constraint.
In order to diagonalize the free theory, we redefine the gravitini as
\begin{equation}
    \psi^{\rm RS}_\mu \defeq \psi_\mu + \frac 18\,\gamma_\mu\lambda \;,\quad \tilde\psi^{\rm RS}_\mu \defeq \tilde\psi_\mu + \frac 18\,\gamma_\mu\tilde\lambda\;,
\end{equation}
and move to the Einstein frame, where the graviton is given by
\begin{equation}
    h_{\mu\nu}^{\rm E} \defeq h_{\mu\nu} + \frac14\,\eta_{\mu\nu}\,\phi \;.
\end{equation}
In this basis, all fields of the NS-NS and NS-R sectors are decoupled and obey standard free field equations (for details, see \cite{Bonezzi:2025anl}). The interpretation of the R-R bispinor $F^{\alpha\tbeta}$, on the other hand, depends on the choice of spinor representations.

\subsubsection*{Type IIB Supergravity}
Let us identify the two spin groups
\begin{equation}
    {\rm Spin}(1,9) \equiv \widetilde{\rm Spin}(1,9)
\end{equation}
and take both copies of spinors in the same Majorana-Weyl representation, thereby identifying $\alpha\equiv\talpha$. 
The R-R bispinor $F^{\alpha\beta}$ can then be expanded by means of the Fierz identity \eqref{Fierz2}, which we rewrite here as
\begin{equation}
    F^{\alpha\beta} = (\gamma^\mu)^{\alpha\beta}F_\mu + \frac1{3!}(\gamma^{\mu\nu\rho})^{\alpha\beta}F_{\mu\nu\rho} + \frac{1}{2\cdot5!}(\gamma^{\mu\nu\rho\sigma\tau})^{\alpha\beta}F_{\mu\nu\rho\sigma\tau} \;,
\end{equation}
yielding a one-form $F_{[1]}$, a three-form $F_{[3]}$ and a self-dual five-form $F_{[5]}$ gauge-invariant field strengths, which are respectively related to a zero-form scalar potential $C_{[0]}$, a two-form potential $C_{[2]}$ and a four-form potential $C_{[4]}$, with self-dual field strength. Indeed, 
from the field equation \eqref{DC_Eom_RR} it follows that such field strengths must obey Bianchi identities and Maxwell's equations given by
\begin{subequations}\begin{align}\label{IIB_RR_Eoms}
    \del_{[\mu} F_{\nu]} &= 0 \;, & \del^\mu F_\mu &= 0 \;, \\[1mm]
    \del_{[\mu} F_{\nu\rho\sigma]} &= 0 \;, & \del^\mu F_{\mu\nu\rho} &= 0 \;, \\[1mm]
    \del_{[\mu} F_{\nu\rho\sigma\tau\omega]} &= 0 \;, & \del^\mu F_{\mu\nu\rho\sigma\tau} &= 0 \;.
\end{align}\end{subequations}
For the case of the five-form, the Bianchi identity implies the Maxwell equation and vice versa, due to the self-duality constraint.

Having identified the two spinor representations, the resulting theory is chiral, with a doublet $(\psi^{{\rm RS}\,\alpha}_\mu,\tilde\psi^{{\rm RS}\,\alpha}_\mu)$ of real Majorana-Weyl gravitini and dilatini $(\lambda_\alpha,\tilde\lambda_\alpha)$. Together with the NS-NS fields and the above Ramond-Ramond forms, they yield the expected spectrum of type IIB supergravity. In this theory, the R-symmmetry group $SO(2)\simeq U(1)$ combines fields from the NS-NS and R-R sectors in a nontrivial way. To present the supersymmetry transformations in a way that is manifestly compatible with the R-symmetry, we first introduce the chiral Weyl spinor parameter $ \upepsilon\defeq \epsilon+i\,\tilde\epsilon$, to which we assign R-charge $+1/2$. We then define the following complex combinations of fields with charge R:
\begin{subequations}\begin{align}
    \uppsi_\mu &\defeq \psi^{\rm RS}_\mu + i\,\tilde\psi^{\rm RS}_\mu\;,\quad\quad {\rm R}=1/2\;,\\
    \uplambda &\defeq \lambda - i\,\tilde\lambda\;,\quad\quad \hspace{9mm}{\rm R}=3/2\;,\\
    A&\defeq\phi-4i\,C\;,\quad\quad\hspace{6mm} {\rm R}=2\;,\\
    A_{\mu\nu}&\defeq B_{\mu\nu}+2i\,C_{\mu\nu}\;,\quad\hspace{3mm} {\rm R}=1\;,
\end{align}\end{subequations}
while the Einstein frame graviton $h^{\rm E}_{\mu\nu}$ and four-form $C_{\mu\nu\rho\sigma}$ are real and have charge zero. Following the various field redefinitions, the supersymmetry transformations \eqref{DC_Glob_SUSY_trans} determined from the double copy can be recast in the following form, which  respects the R-charge of the various fields (see e.g. \cite{Schwarz:1983wa}):
\begin{subequations}\begin{align}
    % Graviton
    \delta_\upepsilon h^{\rm E}_{\mu\nu} &= \tfrac{1}{2}\left(\upepsilon\gamma_{(\mu}\uppsi^*_{\nu)}-\upepsilon^*\gamma_{(\mu}\uppsi_{\nu)}\right) \;, \\
    % Scalar
    \delta_\upepsilon A &= \tfrac{1}{2}\,\upepsilon\uplambda \;, \\
    % 2-form
    \delta_\upepsilon A_{\mu\nu} &=  -\upepsilon\gamma_{[\mu}\uppsi_{\nu]}-\tfrac{1}{8}\,\upepsilon^*\gamma_{\mu\nu}\uplambda \;, \\
    % 4-form
    \delta_\upepsilon C_{\mu\nu\rho\sigma} &= \tfrac{i}{2} \left( \upepsilon^*\gamma_{[\mu\nu\rho}\uppsi_{\sigma]}+\upepsilon\gamma_{[\mu\nu\rho}\uppsi^*_{\sigma]} \right) \;, \\
    % Dilatino
    \delta_\upepsilon \uplambda &=-2i\,\gamma^\mu\del_\mu A\,\upepsilon^*-\tfrac{i}{3}\gamma^{\mu\nu\rho}{\bf F}_{\mu\nu\rho}\,\upepsilon \;, \\
    % Gravitino
    \delta_\upepsilon \uppsi_\mu &=-i\,\del_\nu h^{\rm E}_{\mu\rho}\,\gamma^{\nu\rho}\upepsilon+\tfrac{i}{24}\left(\gamma_\mu{}^{\nu\rho\sigma}{\bf F}_{\nu\rho\sigma}-9\,\gamma^{\nu\rho}{\bf F}_{\mu\nu\rho}\right)\upepsilon^*-\tfrac{1}{2\cdot 5!}\,\gamma^{\nu[5]}\gamma_\mu F_{\nu[5]}\upepsilon \;,
\end{align}\end{subequations}
upon introducing the complex field strength ${\bf F}_{\mu\nu\rho}\defeq 3\,\del_{[\mu}A_{\nu\rho]}$ and recalling that the self-dual field strength for the four form is $F_{\mu_1\dots\mu_5}=5\,\del_{[\mu_1}C_{\mu_2\dots\mu_5]}$.

\subsubsection*{Type IIA Supergravity}
Let us now take the $\widetilde{\rm Spin}(1,9)$ fields to be in the anti-fundamental representation of ${\rm Spin}(1,9)$, so that in general any Majorana-Weyl spinor $\tilde X^\talpha$ in the second copy becomes
\begin{equation}
    \tilde X^\talpha \equiv \tilde X_\alpha
\end{equation}
after the identification of the spin groups. In this way, our new spectrum always contains two copies of its spinors, one in the chiral and the other in the anti-chiral representations.
As a consequence, the physical NS-R fields can be recombined in a Majorana gravitino and dilatino, respectively
\begin{equation}
    \Psi_\mu{} \defeq (\psi^{{\rm RS}\,\alpha}_\mu\,,\tilde\psi^{\rm RS}_{\mu\,\alpha}) \;,\quad
    \uplambda \defeq (\lambda_\alpha\,,\tilde\lambda^\alpha) \;,
\end{equation}
obeying the massless Rarita-Schwinger and Dirac equations.

The R-R bispinor has now the structure $F^\alpha{}_\beta$, and can be expanded using the Fierz identity \eqref{Fierz1}, that is to say
\begin{equation}\label{DC_Fierz_IIA}
    F^\alpha{}_\beta = \delta^\alpha_\beta F + \frac1{2!}(\gamma^{\mu\nu})^\alpha{}_\beta F_{\mu\nu} + \frac1{4!}(\gamma^{\mu\nu\rho\sigma})^\alpha{}_\beta F_{\mu\nu\rho\sigma} \;,
\end{equation}
providing a non-propagating scalar field strength $F_{[0]}$, together with a two-form and a four-form field strengths $F_{[2]}$ and $F_{[4]}$, respectively associated with the existence of a one-form gauge field $C_{[1]}$ and a three-form gauge field $C_{[3]}$. Such field strengths must maintain the gauge invariance of the original bispinor $F^\alpha{}_\beta$ and obey the field equations
\begin{subequations}\begin{align}\label{IIA_RR_Eoms}
    \del_\mu F &= 0 \;, & & \\[1mm]
    \del_{[\mu} F_{\nu\rho]} &= 0 \;, & \del^\mu F_{\mu\nu} &= 0 \;, \\[1mm]
    \del_{[\mu} F_{\nu\rho\sigma\tau]} &= 0 \;, & \del^\mu F_{\mu\nu\rho\sigma} &= 0 \;,
\end{align}\end{subequations}
that are a set of Maxwell's equations and Bianchi identities directly following from plugging the identity \eqref{DC_Fierz_IIA} into the field equation \eqref{DC_Eom_RR}. The above Ramond-Ramond field content, combined with the Majorana spinor and gravitino and the NS-NS sector is exactly the expected spectrum for type IIA supergravity.

%%%%%%%%%%%%%%%%%%%%%%%%%%%%%%%%%%
%%%%%%%%%%%%%%%%%%%%%%%%%%%%%%%%
%% BEGIN CONCLUSIONS %%%%%%%%%%%%%
%%%%%%%%%%%%%%%%%%%%%%%%%%%%%%%%
%%%%%%%%%%%%%%%%%%%%%%%%%%%%%%%%%%
\section{Conclusions}
\label{Sec:Conclusions}

In this paper we have applied the program of using homotopy algebras to derive local, gauge invariant and off-shell double copy relations
to ${\cal N}=1$ super-Yang-Mills theory in $D=10$ in order to obtain type IIA or type IIB supergravity in the same dimension. So far this construction works up to and including quartic couplings \cite{Bonezzi:2022bse}, and the proof of the double copy of the global supersymmetry algebra presented here works up to and including cubic couplings. Specifically, we established the precise map between the double copied theory and standard type II supergravities  at the level of the free theory, but since we  proved that the cubic double copied theory is maximally supersymmetric, it is guaranteed to be equivalent to the corresponding type II supergravity  to this order.

The results presented here should be extended into various  directions. First, it remains as the arguably major outstanding problem to formulate the double copy, taking general gauge theories as the starting point,  to all orders in fields. 
Arguably, this will eventually be achieved by deriving the highly involved BV$_{\infty}^{\square}$ algebra from something much simpler, ideally a strict algebra, say by means of vertex operators  \cite{Bonezzi:2024fhd,DEFquant} and   
homotopy transfer (see \cite{Arvanitakis:2020rrk} for a self-contained review). 
Second, the specific construction presented here suffers from the shortcoming that there is no action principle, as we use a redundant version  of the chain complex 
that is incompatible with a cyclic $L_{\infty}$ structure. To overcome this problem, presumably one has to  introduce new fermionic gauge degrees of freedom. 

Finally, it would be interesting to explore the details for other supersymmetric theories. 
Notably, the superconformal ${\cal N}=(2,0)$ theory in $D=6$ comes to mind, whose double copy was conjectured  
to yield the exotic and so far elusive  $\cN=(4,0)$ supergravity theory of Hull \cite{Hull:2000zn,Borsten:2017jpt}. 
(See \cite{Henneaux:2017xsb,Bertrand:2020nob,Bertrand:2022pyi,Andrianopoli:2022bzr} for partial results on contructing this theory.) 
In particular, since the ${\cal N}=(2,0)$ theory features a superconformal symmetry, as does  the ${\cal N}=4$ theory in four dimensions that we explored in \cite{Bonezzi:2025anl}, 
the obvious question arises what the fate of the conformal symmetries is, given that the double copied supergravities show no signs of them. 
These and other foundational questions we leave for future work.

%%%%%%%%%%%%%%%%%%%%%%%%%%%%%%%%%%
%%%%%%%%%%%%%%%%%%%%%%%%%%%%%%%%
%% ACKNOWLEDGEMENTS %%%%%%%%%%%%%%
%%%%%%%%%%%%%%%%%%%%%%%%%%%%%%%%
%%%%%%%%%%%%%%%%%%%%%%%%%%%%%%%%%%
\section*{Acknowledgments}          % Acknowledgments
We would like to thank L. Andrianopoli, L. Borsten, C. Chiaffrino, G. Itsios, M.F. Kallimani and C. Lavino for discussions and collaborations on closely related topics. 

\noindent The work of R.B.~is funded by the Deutsche Forschungsgemeinschaft (DFG, German Research Foundation) ``Worldline approach to the double copy", Projektnummer 524744955, and G.C.~is funded by the DFG - Projektnummer 417533893/GRK2575 “Rethinking Quantum Field Theory”.

\appendix
%%%%%%%%%%%%%%%%%%%%%%%%%%%%%%%%%%
%%%%%%%%%%%%%%%%%%%%%%%%%%%%%%%%
%% BEGIN APPENDIX 1 %%%%%%%%%%%%%%
%%%%%%%%%%%%%%%%%%%%%%%%%%%%%%%%
%%%%%%%%%%%%%%%%%%%%%%%%%%%%%%%%%%
\section{Spinor Conventions and Identities  }
\label{App:Conventions}
In this paper we use the mostly plus signature convention, such that the ten-dimensional Minkowski metric reads
\begin{equation}
    \eta_{\mu\nu} \defeq (-1\,,1\,,\dots\,,1) \,,
\end{equation}
$\mu,\nu=0,\dots,9$ being the spacetime indices. The totally antisymmetric Levi-Civita tensor $\epsilon_{\mu[10]} \defeq \epsilon_{\mu_1,\dots,\mu_{10}}$ is defined in such a way that
\begin{equation}
    \epsilon^{0123456789} = -\epsilon_{0123456789} = 1\,, \quad \epsilon^{\mu_1,\dots,\mu_{10}}\epsilon_{\mu_1,\dots,\mu_{10}} = -10! \;.
\end{equation}
Let us now go into the details of the spinors in ten dimensions. For more detailed discussions see also \cite{GuillenQuiroz:2016sla,Guillen:2019pnz}.

%%%%%%%%%%%%%%%%%%%%%%%%%%%%%%%%
%%%%%%%%%%%%%%%%%%%%%%%%%%%%%%
%%%%%%%%%%%%%%%%%%%%%%%%%%%%%%%%
\subsection*{Spinors in 10 Dimensions}
Let us consider a ten-dimensional Dirac spinor field $\Psi_{\hat\alpha}$, $\hat\alpha=1,\dots,32$, together with a set of $32\times32$ gamma matrices $\Gamma^\mu$ fulfilling the standard Clifford algebra
\begin{equation}
    \{\Gamma^\mu,\Gamma^\nu\} = 2\,\eta^{\mu\nu} \;.
\end{equation}
In $D = 10$ we can split the Dirac field into its Weyl projections and impose the Majorana condition
\begin{equation}
    \Psi^C \defeq \Psi^T C = \Psi^\dagger\Gamma_0 \;,
\end{equation}
$\Psi^C$ being the charge-conjugate spinor and $C$ the charge-conjugation matrix defined as
\begin{equation}
    C \defeq \begin{pmatrix}
        0 & \delta_\alpha{}^{\dot\beta} \\ -\delta^{\dot\beta}{}_\gamma & 0
    \end{pmatrix}\;,
\end{equation}
so that the Dirac field can be decomposed into its chiral and anti-chiral projections as
\begin{equation}
    \Psi \eqdef \bpm \chi^\alpha \\[1mm] \bar\chi_{\dot\beta} \epm \;,
\end{equation}
with $\bar\chi = \chi^*$ and $\alpha, \dot\alpha = 1, \dots, 16$ so that $\hat\alpha = (\alpha, \dot\alpha)$.

\subsubsection*{Generalized Dirac and Pauli Matrices}
As we want to work with the 16-component spinor $\chi$, we define a generalized set of $16\times16$ Pauli matrices $\gamma^\mu$ via the relation
\begin{equation}
    C\,\Gamma^\mu \eqdef \bpm (\gamma^\mu)_{\alpha\beta} & 0 \\ 0 & (\bar\gamma^\mu)^{\dot{\beta}\dot{\alpha}} \epm \;.
\end{equation}
In $(1,9)$ dimensions, spinors with dotted and undotted indices in the same position (either up or down) transform in the same way under Lorentz transformations \cite{GuillenQuiroz:2016sla}. This allows one to identify the dotted indices with the undotted ones and perform products of $\gamma$ with $\bar\gamma$, showing that they obey the Clifford-like algebra
\begin{equation}
    (\gamma^\mu)_{\alpha\lambda}(\bar\gamma^\nu)^{\lambda\beta} + (\gamma^\nu)_{\alpha\lambda}(\bar\gamma^\mu)^{\lambda\beta} = 2\,\eta^{\mu\nu}\delta_\alpha^\beta \;.
\end{equation}

\noindent Let us now drop the bar symbol and define the antisymmetrizations of the generalized Pauli matrices as
\begin{equation}
    \gamma^{\mu[n]} = \gamma^{\mu_1\dots\mu_n} \defeq \gamma^{[\mu_1}\dots\gamma^{\mu_n]} \;, \quad n\leq D \;,
\end{equation}
which obey the symmetry properties -- with respect to the spinor indices -- listed in Table \eqref{Tab:SymmetryGammaMatrix}, together with the Fierz identity
\begin{equation}
    (\gamma^\mu)_{\alpha(\beta}(\gamma_\mu)_{\lambda\delta)} = 0 \;.
\end{equation}
% % %
\begin{table}[b]
    \centering
    \begin{tabular}{|c|c||c|} 
        \hline
        $n=\text{even}$ & $n=\text{odd}$ & Symmetry \\
        \hline\hline
        0 & 1 & S \\
        2 & 3 & A \\
        4 & 5 & S \\
        6 & 7 & A \\
        8 & 9 & S \\
        10 & -- & A \\
        \hline
    \end{tabular}
    \caption{Symmetry of the gamma matrices $\gamma^{\mu[n]}$ in $D=10$.
    }
    \label{Tab:SymmetryGammaMatrix}
\end{table}
% % %
They also fulfill the following duality relations:
\begin{subequations}\begin{align}
    % % % % 10
    (\gamma^{[10]})_\alpha{}^\beta &= -\epsilon^{[10]}\delta_\alpha{}^\beta \;, &
    (\gamma^{[10]})^\alpha{}_\beta &= +\epsilon^{\mu[10]}\delta^\alpha{}_\beta \;, \\[1mm]
    % % % % 9
    (\gamma^{\mu[9]})_{\alpha\beta} &= -\epsilon^{\mu[9]\nu[1]}(\gamma_{\nu[1]})_{\alpha\beta} \;, &
    (\gamma^{\mu[9]})^{\alpha\beta} &= +\epsilon^{\mu[9]\nu[1]}(\gamma_{\nu[1]})^{\alpha\beta} \;, \\[1mm]
    % % % % 8
    (\gamma^{\mu[8]})_\alpha{}^\beta &= +\frac12\epsilon^{\mu[8]\nu[2]}(\gamma_{\nu[2]})_\alpha{}^\beta \;, &
    (\gamma^{\mu[8]})^\alpha{}_\beta &= -\frac12\epsilon^{\mu[8]\nu[2]}(\gamma_{\nu[2]})^\alpha{}_\beta \;, \\[1mm]
    % % % % 7
    (\gamma^{\mu[7]})_{\alpha\beta} &= +\frac1{3!}\epsilon^{\mu[7]\nu[3]}(\gamma_{\nu[3]})_{\alpha\beta} \;, &
    (\gamma^{\mu[7]})^{\alpha\beta} &= -\frac1{3!}\epsilon^{\mu[7]\nu[3]}(\gamma_{\nu[3]})^{\alpha\beta} \;, \\[1mm]
    % % % % 6
    (\gamma^{\mu[6]})_\alpha{}^\beta &= -\frac1{4!}\epsilon^{\mu[6]\nu[4]}(\gamma_{\nu[4]})_\alpha{}^\beta \;, &
    (\gamma^{\mu[6]})^\alpha{}_\beta &= +\frac1{4!}\epsilon^{\mu[6]\nu[4]}(\gamma_{\nu[4]})^\alpha{}_\beta \;, \\[1mm]
    % % % % 5
    (\gamma^{\mu[5]})_{\alpha\beta} &= -\frac1{5!}\epsilon^{\mu[5]\nu[5]}(\gamma_{\nu[5]})_{\alpha\beta} \;, &
    (\gamma^{\mu[5]})^{\alpha\beta} &= +\frac1{5!}\epsilon^{\mu[5]\nu[5]}(\gamma_{\nu[5]})^{\alpha\beta} \;.
\end{align}\end{subequations}
Using the Weyl projections and the Pauli matrices $\gamma^\mu$, one can write down two more major Fierz identities
\begin{subequations}\begin{align}
    \xi_\alpha\chi^\beta &= \frac1{16}\delta^\beta_\alpha(\xi\chi) + \frac1{2!}\frac1{16}(\gamma^{\mu\nu})_\alpha{}^\beta(\xi\gamma_{\mu\nu}\chi) + \frac1{4!}\frac1{16}(\gamma^{\mu\nu\rho\sigma})_\alpha{}^\beta(\xi\gamma_{\mu\nu\rho\sigma}\chi) \;, \label{Fierz1}\\[1mm]
    % % %
    \xi^\alpha\chi^\beta &= \frac1{16}(\gamma^\mu)^{\alpha\beta}(\xi\gamma_\mu\chi) + \frac1{3!}\frac1{16}(\gamma^{\mu\nu\rho})^{\alpha\beta}(\xi\gamma_{\mu\nu\rho}\chi) + \frac1{5!}\frac1{32}(\gamma^{\mu\nu\rho\sigma\tau})^{\alpha\beta}(\xi\gamma_{\mu\nu\rho\sigma\tau}\chi) \;, \label{Fierz2}
\end{align}\end{subequations}
which hold in our ten-dimensional case.

\subsubsection*{Useful Properties}
Here we provide a list of useful properties of the $\gamma$-matrices. The multiplication of arbitrary $\gamma^{\mu_1\dots\mu_p}$ by a single $\gamma^\mu$ yields
\begin{subequations}\begin{align}
    \gamma^\mu\gamma^\nu &= \gamma^{\mu\nu} + \eta^{\mu\nu} \;, &
    \gamma^\mu\gamma_\mu &= D \;, \\[1mm]
    \gamma^{\mu\nu}\gamma^\rho &= \gamma^{\mu\nu\rho} + 2\,\gamma^{[\mu}\eta^{\nu]\rho} \;, &
    \gamma^{\mu\nu}\gamma_\nu &= (D-1)\,\gamma^\mu \;, \\[1mm]
    \gamma^{\mu\nu\rho}\gamma^\sigma &= \gamma^{\mu\nu\rho\sigma} + 3\,\gamma^{[\mu\nu}\eta^{\rho]\sigma} \;, &
    \gamma^{\mu\nu\rho}\gamma_\rho &= (D-2)\,\gamma^{\mu\nu} \;, \\[1mm]
    \gamma^{\mu\nu\rho\sigma}\gamma^\tau &= \gamma^{\mu\nu\rho\sigma\tau} + 4\,\gamma^{[\mu\nu\rho}\eta^{\sigma]\tau} \;, &
    \gamma^{\mu\nu\rho\sigma}\gamma_\sigma &= (D-3)\,\gamma^{\mu\nu\rho} \;, \\[1mm]
    \gamma^{\mu\nu\rho\sigma\tau}\gamma^\lambda &= \gamma^{\mu\nu\rho\sigma\tau\lambda} + 5\,\gamma^{[\mu\nu\rho\sigma}\eta^{\tau]\lambda} \;, &
    \gamma^{\mu\nu\rho\sigma\tau}\gamma_\tau &= (D-4)\,\gamma^{\mu\nu\rho\sigma} \;, \\
    &\hspace{2mm}\vdots & &\hspace{2mm}\vdots \nonumber \\
    \gamma^{\mu_1\dots\mu_p}\gamma^\nu &= \gamma^{\mu_1\dots\mu_p\nu} + p\,\gamma^{[\mu_1\dots\mu_{p-1}}\eta^{\mu_p]\nu} \;, &
    \gamma^{\mu_1\dots\mu_p\nu}\gamma_{\nu} &= (D-p)\,\gamma^{\mu_1\dots\mu_p} \;,
\end{align}\end{subequations}
where we kept $10=D$ for the traces.
\noindent Multiplying the above relations by a further $\gamma$-matrix and antisymmetrizing one obtains
\begin{subequations}\begin{align}
    \gamma_{\mu\nu}\gamma^{\rho\sigma} &= \gamma_{\mu\nu}{}^{\rho\sigma} - 4\,\gamma_{[\mu}{}^{[\rho}\delta_{\nu]}^{\sigma]} - 2\,\delta^{\rho\sigma}_{\mu\nu} \;, \\[1mm]
    \gamma_{\mu\nu\rho}\gamma^{\sigma\tau} &= \gamma_{\mu\nu\rho}{}^{\sigma\tau} - 6\,\gamma_{[\mu\nu}{}^{[\sigma}\delta_{\rho]}^{\tau]} - 6\,\gamma_{[\mu}\delta^{\sigma\tau}_{\nu\rho]} \;, \\[1mm]
    \gamma_{\mu\nu\rho\sigma}\gamma^{\tau\omega} &= \gamma_{\mu\nu\rho\sigma}{}^{\tau\omega} - 8\,\gamma_{[\mu\nu\rho}{}^{[\tau}\delta_{\sigma]}^{\omega]} - 12\,\gamma_{[\mu\nu}\delta^{\tau\omega}_{\rho\sigma]} \;, \\[1mm]
    \gamma_{\mu\nu\rho\sigma\tau}\gamma^{\omega\lambda} &= \gamma_{\mu\nu\rho\sigma\tau}{}^{\omega\lambda} - 10\,\gamma_{[\mu\nu\rho\sigma}{}^{[\omega}\delta_{\tau]}^{\lambda]} - 20\,\gamma_{[\mu\nu\rho}\delta^{\omega\lambda}_{\sigma\tau]} \;, \\
    &\hspace{2mm}\vdots \nonumber \\
    \gamma_{\mu_1\dots\mu_p}\gamma^{\nu\rho} &= \gamma_{\mu_1\dots\mu_p}{}^{\nu\rho} - 2p\,\gamma_{[\mu_1\dots\mu_{p-1}}{}^{[\nu}\delta_{\mu_p]}^{\rho]} - p(p-1)\gamma_{[\mu_1\dots\mu_{p-2}}\delta_{\mu_{p-1}\mu_p]}^{\nu\rho} \;.
\end{align}\end{subequations}
Finally, other partial traces are given by
\begin{subequations}\begin{align}
    \gamma_{\mu\rho}\gamma^{\rho\nu} &= (D-1)\delta_\mu^\nu + (D-2)\gamma_\mu{}^\nu \;, \\[1mm]
    \gamma_{\mu\nu}\gamma^{\mu\nu} &= (1-D)D \;, \\[1mm]
    \gamma^\mu\gamma^\nu\gamma_\mu &= (2-D)\gamma^\nu \;, \\[1mm]
    \gamma^\mu\gamma^\nu\gamma^\rho\gamma_\mu &= (D-4)\gamma^{\nu\rho} + D\eta^{\nu\rho} \;, \\[1mm]
    \gamma_{\mu\nu}\gamma_\rho\gamma^{\mu\nu} &= -(D-1)(D-4)\gamma_\rho \;, \\[1mm]
    \gamma_{\rho\sigma}\gamma^{\mu\nu}\gamma^{\rho\sigma} &= -[(D-7)(D-2)+2]\gamma^{\mu\nu} \;, \\[1mm]
    \gamma^\nu\gamma_{\mu[p]}\gamma_\nu &= -(D+(-1)^p\,2p)\gamma_{\mu[p]} \;, \\[1mm]
    \gamma^\nu\gamma_{\mu[5]}\gamma_\nu &= 0 \quad \text{for} \quad D=10\,,\quad p=5 \;.
\end{align}\end{subequations}

%%%%%%%%%%%%%%%%%%%%%%%%%%%%%%%%%%
%%%%%%%%%%%%%%%%%%%%%%%%%%%%%%%%
%% BIBLIOGRAPHY %%%%%%%%%%%%%%%%%%
%%%%%%%%%%%%%%%%%%%%%%%%%%%%%%%%
%%%%%%%%%%%%%%%%%%%%%%%%%%%%%%%%%%
\providecommand{\href}[2]{#2}\begingroup\raggedright\endgroup

\end{document}